\documentstyle[aps,pre,floats,epsf,eqsecnum,psfig]{revtex}

\newcommand{\vect}[1]{ {\bf #1} }
\newcommand{\vectg}[1]{\mbox{\boldmath $#1$}}
\newcommand{\vectd}[2]{\mbox{\boldmath $#1$}{\bf #2}}
\def \e {\mbox{e}}
\begin{document}

\title{Instanton Calculus in Shell Models of Turbulence}
\author{Isabelle Daumont$^1,2$, Thierry Dombre$^1$ and Jean-Louis Gilson$^{1}$}
\address{$^1$ Centre de Recherches sur les Tr\`es Basses Temp\'eratures-CNRS,
 Laboratoire conventionn\'e avec l'Universit\'e Joseph Fourier,
BP166, 38042 Grenoble C\'edex~9, France\\
$^2$ Ecole normale sup\'erieure de Lyon, Laboratoire de Physique,
 46 all\'ee d'Italie, 69364 Lyon C\'edex 07}
\date{}
\maketitle
\begin{abstract}
It has been shown recently that intermittency of the Gledzer Ohkitani 
Yamada (GOY) shell model of turbulence has to be related to singular
 structures whose dynamics in the inertial range includes interactions
with a background of fluctuations. In this paper we propose a statistical 
theory of these objects by modelling the incoherent background as a Gaussian
 white-noise forcing of small strength $\Gamma $. A general scheme is developed
for constructing instantons in spatially discrete dynamical systems and
the Cram\'er function governing the probability distribution of effective
 singularities of exponent $z$ is computed up to first order in a semiclassical
expansion in powers of $\Gamma $. The resulting predictions are compared
with the statistics of coherent structures deduced from full simulations of
 the GOY model at very high Reynolds numbers. 
\end{abstract}

\section{Introduction}
Are structures (sheets or filaments of vorticity) a vital ingredient of
intermittency in 3D-incom\-pres\-sible turbulence? 
To date, this 
important question remains open \cite{3D}, and an answer starting from first 
principle, i.e., from a controlled approximation to the Navier-Stokes 
equations, seems over the horizon. The new understanding of the anomalous 
scaling in the Kraichnan's model of passive advection \cite{PS}, based on 
the identification of zero modes in
the homogeneous Hopf equations for equal-time correlators, has in particular
strengthened the belief that field-theoretical methods would eventually be 
able to capture the full statistics of turbulent flows without 
an explicit account of structures.

Interestingly enough, the relative interplay between coherent ordered
structures and incoherent turbulent fluctuations turns out to be a subtle
matter already in the restricted framework of the so-called shell models of
 turbulence \cite{SM}. It was noticed very soon \cite{SN} that elementary 
bricks
 of intermittency in those deterministic 1D-cascade models could be pulses
or bursts of activity growing in an almost self-similar way as they move
from large to small scales. However, genuine dynamically stable self-similar
 solutions of the equations of motion in the inertial range display an unique
 scaling exponent (to be denoted below as $z_{0}$), provided they are
 localized in $\vect{k}$-space (which, in the shell model approach, reduces
 to a discrete set of wave numbers $k_{n}=2^{(n-1)}$, where the shell index
 $n$ goes from 1 to $\infty$). Furthermore, the exponent $z_{0}$, giving the
 logarithmic slope of the velocity gradient spectrum left in the trail of the
 pulse, happens to be rather close to the Kolmogorov value $2/3$
 ($z_{0}=0.72$) in the case of the Gledzer-Okhitani-Yamada (GOY) model, in
 the range of parameters where it reproduces at best the multiscaling
 properties of real turbulent flows.

In Ref.\cite{GD97}, the role played by the interaction of pulses
with the rest of the flow in producing more singular events was unravelled,
and a two-fluid picture was introduced, where coherent structures form in
and propagate into a featureless random background. Our goal in this paper
is to elevate this still rather qualitative proposal to the rank
 of a semi-quantitative theory and to test its predictive power about
intermittency in the GOY model. We shall assume that turbulent fluctuations
 on the shells downstream the pulse, i.e., small scales, act on the coherent
part of the flow as a random, white-in-time, Gaussian forcing and ask
whether the inviscid stochastic extension of the GOY model obtained in this
 way is able 
to reproduce the statistics of strong deviations of the full turbulent system
in the inertial range. There is {\em a priori} quite a lot of freedom in the
parameterisation of the forcing. Therefore, in order to keep things as
simple as possible, we bind ourselves to use a single
adjustable parameter (hereafter noted $\Gamma $), which measures the level
of noise. We consider the semiclassical limit $\Gamma \ll 1$ of these 
systems and study the statistics of singular structures appearing in this
regime. 

Semiclassical (or instanton) techniques are well suited to capture large and
 rare excursions of fluctuating fields \cite{LIP}. As such, they have
 gained recently a renewal of interest in the field of turbulence and have
 already led to noteworthy results in the context of Burger's turbulence
 \cite{GM96,BFKL97}, and of the Kraichnan's model of passive scalar
 advection \cite{C97,BL98}. One usually starts from a path integral
 representation of high order structure functions and uses a saddle point
 approximation to determine the coupled field-force configurations
 contributing mostly to those quantities.   
 The nature of the statistical object to be computed imposes precise
 boundary conditions on the physical field and the random force
 (respectively at large and small scales, where the cascade processes start
and end). Instantons, which in the inertial range often reduce to a
 self-similar collapse along some spatial dimensions, are eventually
 selected by a delicate matching procedure at the two boundaries. 
In shell models we are dealing with an intrinsically discrete lattice of
 logarithmic scales. As a consequence, the analytic computation of instantons
 is completely out of reach in the inertial range, not to speak about
 the matching on both sides of the cascade. To circumvent this difficulty,
we shall focus on the probability distribution function (pdf) of scaling
 exponents after $n$ cascade steps, $P_{n}(z)$, and argue that, in the
 semiclassical limit $\Gamma \ll 1$, this pdf builds up from the
 neighborhood of a single self-similar instanton (of scaling exponent $z$)
 that dynamic stability considerations will help us to construct numerically.
 In order to get non trivial physics, it turns out to be necessary to
 perform the semiclassical expansion of
 $-\lim_{n\rightarrow \infty}\frac{1}{n}\ln P_{n}(z)$
 (the rate of rarefaction of singularities of scaling exponent $z$ in the
 multifractal picture) up to next to leading order in powers of $\Gamma $.
This can be achieved via a summation over quadratic fluctuations around
 the instantons, once the proper set of boundary conditions for the 
corresponding trajectories in configuration space has been defined. 
We shall show in details how to carry out this program and end up with a
 prediction for $P_{n}(z)$ lending itself to a straight confrontation
with the pdf of effective scaling exponents of coherents events that can
be extracted from simulations of the GOY model at very high Reynolds numbers.
Although our interest lies primarily in gaining a better understanding
of intermittency in the framework of shell models of turbulence, the
emphasis will be put in this paper on the technical aspects of the method
that we had to develop for computing instantons. We believe that this 
method is general enough to find applications in other contexts or
physical problems, like for instance the motion of complex objects or
excitations on 1D-lattices in the presence of a co-moving random
environment.

The paper is organized as follows. In Section \ref{def}, we define the 
stochastic extensions of the GOY model that we shall study. In Section
 \ref{MSRformalism} the equations of motion for instantons will be derived 
 using the well-known Martin Siggia Rose path integral representation of 
probability distribution functions for stochastic dynamical systems. Section
 \ref{ExtSol} is devoted to the computation of self-similar extremal
trajectories, with the theoretical considerations underlying the solution
 explained in Subsection \ref{ExtSoltheory} and its practical implementing,
 together with the results, exposed in Subsection \ref{ExtSolimplementing}.
The important effect of quadratic fluctuations and the rather heavy
formal work behind their computation are discussed in Section \ref{fluct}.
The comparison of the results issuing from the instanton approach with
 numerical data on the statistics of coherent structures in the genuine
 GOY model is given in Section \ref{discussion}. We conclude in Section
 \ref{conclusion}, in particular as to the relevance of a two-fluid
 description of intermittency in shell models of turbulence.

\section{Definition of the Stochastic Dynamical System}
\label{def}

Equations of motion for the GOY model in the inertial range 
read~:
\begin{equation}
\frac{db_{n}}{dt}  =  Q^{2}(1-\epsilon)b^{*}_{n-2}b^{*}_{n-1} 
+ \epsilon b^{*}_{n-1}b^{*}_{n+1} - Q^{-2}b^{*}_{n+1}b^{*}_{n+2},
\label{GOYmod}
\end{equation}
where the complex variable $b_{n}=k_{n}u_{n}$ should be understood as the
Fourier component of the gradient velocity field at wavenumber $k_{n}=Q^{n}$
and the integer $n$ runs from $0$ to $+\infty$. Throughout this paper,
usual values of parameters $\epsilon=0.5$ and $Q=2$ will be assumed. 
It is convenient to cast (\ref{GOYmod}) in a vectorial form
\begin{equation}
\frac{d\vect{b}}{dt}= \vect{N}[\vect{b}],
\label{GOYvect}
\end{equation}
where the infinite-dimensional vector $\vect{b}$ is built up
from the $b_n$'s, while the $n^{\mbox{th}}$ component of the nonlinear
kernel $\vect{N}[\vect{b}]$ is given by the right hand side of (\ref{GOYmod}).
It is worth noting at this point that $\vect{b}^{*}\cdot\vect{b}=
\sum_{n=0}^{\infty} |b_{n}|^{2}$ plays dimensionally the role of
enstrophy in real flow and that the inverse square
root of this quantity sets the order of magnitude of the smallest
time scale on the shell lattice. \\

Since quadratic nonlinearities lead generically to finite time singularities,
it is very useful to introduce a desingularizing time variable
$\tau $ related to the physical time $t$ by the differential law
\begin{equation}
\frac{d\tau }{dt} =(\vect{b}^{*}\cdot\vect{b})^{1/2}.
\label{def:tau}
\end{equation}
This turns (\ref{GOYvect}) into
\begin{equation}
\frac{d\vect{b}}{d\tau }=
 \frac{\vect{N}[\vect{b}]}{(\vect{b}^{*}\cdot\vect{b})^{1/2}},
\label{GOYtau}
\end{equation}
where both sides of the equation have the same scaling dimension in the
field $\vect{b}$, which shows that an infinite ``time'' is now required to 
form a singularity by travelling across the whole shell axis.

>From previous work \cite{DG98}, we know that every initial condition of finite
 enstrophy, when evolving under dynamics (\ref{GOYtau}), eventually organizes itself 
in a soliton-like pulse, moving from large to small scales at a
 constant speed with an exponential growth of its amplitude. The asymptotic
state is unique, up to trivial phase symmetries of the GOY model \cite{BBP93},
time translations or multiplicative rescaling of the field $\vect{b}$, which
all leave the equation of motion (\ref{GOYtau}) invariant. We may restrict
our attention without loss of generality to the case where the phase pattern
 along the shell axis does not break into a three-sublattice structure. The
asymptotic Floquet state, to be noted henceforth $\vect{b}^{0}(\tau )$, is then purely 
real and such that~:
\begin{equation}
b^{0}_{n+1}(\tau +T_{0})=\exp (A_{0}T_{0}) b^{0}_{n}(\tau ).
\label{def:TandA_0}
\end{equation}
The period $T_{0}$ is the ``time'' needed for the center of the
pulse to go from shell $n$ to shell $n+1$, while the (positive) Lyapunov exponent 
$A_{0}$ controls its growth. Both quantities $T_{0}$ and $A_{0}$, are dynamically selected
 in an unique way. The scaling exponent $z_{0}$ associated
to the pulse (fixing in particular the logarithmic slope of the spectrum
left in its trail) can be extracted from the identity 
$Q^{z_{0}}= \exp (A_{0}T_{0})$. Its value turns out to be 0.72 in the case of the GOY model 
for the choice of parameters stated before.\\


We turn now to the stochastic models, that we are interested in solving by the
instanton method. Their physical motivation has been explained in
Ref~. \cite{GD97}~: we assume that pulses parameterize adequately
singular (and temporally coherent) structures in shell models but that
the deterministic dynamics (\ref{GOYvect}) should be enlarged towards
a stochastic one, in order to describe the interaction of a given 
pulse with incoherent fluctuations produced by the relaxation of
the trails left by its predecessors. We are therefore led to consider
the following extension of the original inviscid GOY model~:
\begin{equation}
\label{def:syststoch}
\frac{d\vect{b}}{dt}= \vect{N}[\vect{b}] + \sqrt{\Gamma}(\vect{b}^{*}\cdot\vect{b})^{3/4}
 B[\vect{C}] \vectg{\eta}\,,
\end{equation}
where $\vectg{\eta}$ is a Gaussian noise, delta-correlated in
time and shell index, whose correlations read~:
\begin{equation}
\label{def:correl1}
\langle \eta _n^{*}(t) \eta_{n'}(t')\rangle =\delta _{nn'} \delta (t-t')\,.
\end{equation}
The various factors coming in front of $\vectg{\eta}$ in (\ref{def:syststoch})
 have the following 
meaning~: the number $\Gamma $ fixes the relative strength of incoherent
fluctuations with respect to coherent ones and we shall be interested
in the semiclassical limit of small $\Gamma $ amenable to semi-analytic
treatment. As will be clearer in a while, the overall scale factor 
 $(\vect{b}^{*}\cdot\vect{b})^{3/4}$ is there to keep noise relevant all
along the cascade, thereby preserving
scale invariance. Finally the matrix $B[\vect{C}]$,
of zero scaling dimension in the field $\vect{b}$ since it depends only
on the unit vector $\vect{C}=\frac{\vect{b}}{\sqrt{(\vect{b}^{*}\cdot\vect{b})}}$,
may be used either to introduce spatial correlations of noise 
(along the shell axis) or to localize its action with respect to the
 instantaneous
 position of the pulse. 
Although the formalism to be developed in this paper can deal with the most 
general situation, we restricted ourselves in practical investigations to diagonal 
matrices $B[\vect{C}]$, just playing with the degree of localization 
of noise. Results will be presented for three rather emblematic choices of $B$~:
(i) $B_{nn}=1$, which describes a completely delocalized noise; (ii) 
$B_{nn}=C^{*}_{n-2}C^{*}_{n-1}$ which keeps some flavour of the
original GOY dynamics and makes noise active just at the leading
edge of the pulse; and finally (iii) $B_{nn}=|C_{n-5}|^{2}+|C_{n-4}|^{2}$, 
which removes the action of noise further away from the center
of the pulse. We must emphasize that these particular choices were not
dictated by rigorous considerations on the underlying
dynamics of the GOY model, but rather used to scan the variety of  
behaviours which may be expected from such stochastic
dynamical systems. It should be noted that the strucutre of the matrix
$B$ is not constrained by any conservation law, since the coherent part
of the flow does not form anymore a closed system, even in the inertial range,
in our two-fluid description. Finally, to simplify the following analysis, we 
are going to restrict the fields $\vect{b}$ and $\vectg{\eta}$ in 
(\ref{def:syststoch}) and (\ref{def:correl1}) to being real-valued vectors
and neglect the effect of imaginary fluctuations. This is certainly not a
serious restriction as for the instantons themselves, which are expected to
be, like the self-similar deterministic solution described above, purely real,
 up to trivial phase symmetries of the GOY model. It can also be remarked 
that the model (ii) (which will be found later on to give the more convincing
results) does not require a complex noise, since the phase degrees of freedom
have already been incorporated in the definition of the matrix B in that case. \\

While the deterministic dynamics (\ref{GOYvect}) selects a single self-similar
solution blowing up in finite time with scaling exponent $z_{0}$, 
the presence of noise in (\ref{def:syststoch}) allows for a continuum of 
scaling exponents, even in the manifold of normalizable fields $\vect{b}$.
In the small noise (or semiclassical) limit $\Gamma \ll 1$, 
the probability density of developing an effective growth exponent 
$z$ after $n\gg 1$ cascade steps will take the form ~:
\begin{equation}
\label{pred_Pn}
P_n(z){\sim}\sqrt{n}\, \exp \left[-n\left({s_0(z)\over\Gamma}
+ s_1(z)\right)\right], 
\end{equation}
where $s_0(z)$ is the action per unit cascade step of the self-similar extremalsolution of scaling exponent $z$ of optimal bare Gaussian weight
 (or instanton), and
 $s_1(z)$ measures, to lowest order in $\Gamma $, how the basin of attraction
of the instanton in phase space evolves with the number of cascade steps.
 Note that the argument of the exponential in (\ref{pred_Pn}),
 $-(\frac{s_{0}(z)}{\Gamma}+s_{1}(z))$,
is nothing but the Cram\'er function introduced in the theory of
large deviations, which governs the rate of rarefaction of singularities in the
multifractal picture \cite{UF}. We will show in this paper
how to compute in a clean way both quantities $ s_0(z)$ and $s_1(z)$.
Before doing this, we must carefully handle problems related
to the time discretization of the stochastic equation (\ref{def:syststoch})
since a consistent treatment of them is necessary to get the right
expression of the first correction $s_{1}(z)$. We shall adopt the view
that the initial stochastic equation (\ref{def:syststoch}) is to be understood
in the Stratonovich sense \cite{G82}.
 However, in the path integral formulation of stochastic dynamical systems 
that we shall heavily use in the following, it is 
much simpler to work with the Ito prescription which, in the
 limit of small time steps, amounts integrating (\ref{def:syststoch})
within a basic Euler scheme with all $\vect{b}$-dependent quantities in the 
r.h.s. estimated at the prepoint. When switching to the Ito discretization
recipe, the stochastic equation has to be changed into~:
\begin{equation}
\label{Ito_timet}
\frac{d\vect{b}}{dt}= \vect{N}_{\Gamma }[\vect{b}] +  \sqrt{\Gamma}
(\vect{b}\cdot\vect{b})^{3/4} B[\vect{C}] \vectg{\eta}\,,
\end{equation}
where the new kernel $\vect{N}_{\Gamma }[\vect{b}]$ differs from $\vect{N}[\vect{b}]$
 by the addition of the so-called Ito drift-term. We give,
for the sake of completeness, the resulting expression of the $n^{th}$
component of $\vect{N}_{\Gamma}$~:
\begin{equation}
\label{Itokernel}
N_{\Gamma\,n}[\vect{b}]=N_{n}[\vect{b}]+\frac{1}{2}\Gamma \,
\frac{\partial }{\partial b_{k}}[(\vect{b}\cdot\vect{b})^{3/4}B_{nj}]
(\vect{b}\cdot\vect{b})^{3/4}B_{kj}\,.
\end{equation}
At this point, we may write down the discrete analog of
(\ref{def:tau}) as $\Delta \tau_{i}=\tau_{i+1}-\tau_{i}
=(\vect{b}_{i}\cdot\vect{b}_{i})^{1/2}(t_{i+1}-t_{i})$ (where $i$ is the
time index) and redefine the noise as $\vectg{\eta}_{i}\rightarrow 
\vectg{\xi}_{i}=\sqrt{\Gamma}(\vect{b}_{i}\cdot\vect{b}_{i})^{-1/4}\,
\vectg{\eta}_{i}$. This
leads to the following stochastic extension of (\ref{GOYtau}) which will be
the starting point of our formal analysis~:
\begin{equation}
\label{Ito_tau}
\frac{d\vect{b}}{d\tau }= \frac{\vect{N}_{\Gamma }[\vect{b}]}
{(\vect{b}\cdot\vect{b})^{1/2}} + (\vect{b}\cdot\vect{b})^{1/2} 
B[\vect{C}] \vectg{\xi}\,,
\end{equation}
with
\begin{equation}
\label{def:correl2}
\langle \xi _n(\tau ) \xi_{n'}(\tau ')\rangle =\Gamma \delta _{nn'}
 \delta (\tau -\tau ')\,.
\end{equation}

\section{Extremal trajectories from path integral formulation }
\label{MSRformalism}
 Statistics of classical fields in the presence of random forces can be
 examined with the help of field theoretical techniques formulated in 
\cite{MSRandCo}. In particular, the probability to go from point
$\vect{b}_{in}$ at time $\tau =0$ to point $\vect{b}_{f}$ at time $\tau_{f}$
may be written as a path integral~:
\begin{equation}
\label{pathint}
P(\vect{b}_{in},0;\vect{b}_{f}, \tau_{f})=\int {\cal D} \vect{b}\,
{\cal D} \vect{p}\,\exp -S[\vect{b}, \vect{p}]\,,
\end{equation}
where $S[\vect{b},\vect{p}]$ is an effective action to be defined below, 
$\vect{p}$ an auxiliary field conjugated to the physical one $\vect{b}$ and
${\cal D} \vect{b}\,{\cal D} \vect{p}$ stands for~:
\begin{equation}
\frac{d\vect{p}_{0}}{(2\pi )^{d}}\,\prod _{i=1}^{i=N-1}\frac{d\vect{b}_{i}
d\vect{p}_{i}}{(2\pi )^{d}}\,.
\end{equation}
In the last equation, the time interval $\tau_{f}$ was divided into $N$
 subintervals
of length $\Delta \tau=\frac{\tau_{f}}{N}$ (with $\vect{b}_{in}=\vect{b}_{0}$ 
and $\vect{b}_{f}=\vect{b}_{N}$) and the number of shells was set
to a finite value $d$, in order to give a clear meaning to the measure. For
the problem of interest (\ref{Ito_tau}), the effective action $S$ takes the
form~:
\begin{equation}
\label{MSRaction1}
S[\vect{b},\vect{p}]=\sum_{i=0}^{i=N-1} i\vect{p}_{i}.\left(\vect{b}_{i+1}-
\vect{b}_{i}-\Delta \tau \frac{\vect{N}_{\Gamma}[\vect{b}_{i}]}
{(\vect{b}_{i}\cdot\vect{b}_{i})^{1/2}}\right)+\frac{\Gamma }{2}
(\vect{b}_{i}\cdot\vect{b}_{i})\,
\vect{p}_{i}\cdot B[\vect{b}_{i}]\,^{t}\! B[\vect{b}_{i}]\vect{p}_{i}\,,
\end{equation}
or, in the continuum limit~:
\begin{equation}
\label{MSRaction2}
S[\vect{b},\vect{p}]=\int _{0}^{\tau_{f}}\,d\tau \,
i\vect{p}.\left(\frac{d\vect{b}}{d\tau }-\frac{\vect{N}_{\Gamma}[\vect{b}]}
{(\vect{b}\cdot\vect{b})^{1/2}}\right)+\frac{\Gamma }{2}(\vect{b}\cdot\vect{b})\,
\vect{p}\cdot B[\vect{b}]\,^{t}\! B[\vect{b}]\vect{p}\,.
\end{equation}
The last term in (\ref{MSRaction2}), quadratic in $\vect{p}$, appears as a result
of averaging over the Gaussian noise $\vectg{\xi }$, while the first one, 
linear in $\vect{p}$, would still be there in the absence of noise as a formal
 way of enforcing the deterministic equation of motion of $\vect{b}$. $S[\vect{b},\vect{p}]$ will be referred to in the following as the 
Martin-Siggia-Rose (MSR) action. \\

Rescaling the auxiliary field $\vect{p}$ as $\vect{p}'/\Gamma $
puts an overall large factor $1/\Gamma $ in front of the effective
action and opens the way to a saddle point approximation to the path integral (\ref{pathint}). Extremization of the 
action with respect to the configurations of both fields $\vect{b}$ and
$\vect{p}$ between times $0$ and $\tau _{f}$, for fixed endpoints, leads
in a straightforward way to the following set of coupled equations
defining extremal trajectories~:
\begin{eqnarray}
\label{extremalesMSR1}
\frac{d\vect{b}}{d\tau}& = &\frac{\vect{N}_{\Gamma }[\vect{b}]}
{(\vect{b}\cdot\vect{b})^{1/2}} + (\vect{b}\cdot\vect{b}) 
B\,^{t}\! B \vectg{\theta }\,,\\
\frac{d\vectg{\theta }}{d\tau }& = & - ^{t}\!{\cal M}\,\vectg{\theta }
-\frac{1}{2}{\partial }_{\vect{b}}[(\vect{b}\cdot\vect{b})\,
\vectg{\theta }\cdot B\,^{t}\!B\vectg{\theta }]\,.
\label{extremalesMSR2}
\end{eqnarray}
In the above equation, we set $\vect{p}'=-i\,\vectg{\theta}$ and 
${\cal M}$ is the Jacobian matrix of the kernel 
$\frac{\vect{N}_{\Gamma }[\vect{b}]}{(\vect{b}\cdot\vect{b})^{1/2}}$~:
${\cal M}=\displaystyle{\frac{\partial _{\vect{b}} 
\vect{N}_{\Gamma }[\vect{b}]}{(\vect{b}\cdot\vect{b})^{1/2}}-
\frac{\vect{N}_{\Gamma}[\vect{b}] \otimes
\vect{b}}{(\vect{b}\cdot\vect{b})^{3/2}}}$.
As usual, equations (\ref{extremalesMSR1}) and 
(\ref{extremalesMSR2}) inherit a canonical structure
\begin{eqnarray}
\label{canonical1}
\frac{d\vect{b}}{d\tau}& = &\frac{\partial {\cal H}}
{\partial \vectg{\theta}}\,,\\
\frac{d\vectg{\theta}}{d\tau}& = &-\frac{\partial {\cal H}}
{\partial \vect{b}}\,,
\label{canonical2}
\end{eqnarray}
where the Hamiltonian $\cal H$ reads
\begin{equation}
\label{Hamiltonian}
{\cal H}=\frac{\vectg{\theta}\cdot\vect{N}_{\Gamma}[\vect{b}]}
{(\vect{b}\cdot\vect{b})^{1/2}}+\frac{1}{2}(\vect{b}\cdot\vect{b})
(\vectg{\theta}\cdot B\,^{t}\!B\vectg{\theta})\,.
\end{equation}
Since $\cal H$ is not explicitly time-dependent, we conclude that its value,
to be called the pseudo-energy in the sequel, is conserved along any
 extremal trajectory. The action $S[\vect{b}, \vect{\theta}]$ may be 
rewritten in terms of $\cal H$ as
\begin{equation}
S[\vect{b},\vectg{\theta}]=\int _{0}^{\tau_{f}}d\tau \left(\vectg{\theta}.
\frac{d\vect{b}}{d\tau }-{\cal H}\right)\,,
\end{equation}
from which it is seen that the further requirement that the trajectory be
extremal with respect to time reparametrization leads to the condition
of vanishing pseudo-energy ${\cal H}=0$.
Noting that each term of $\cal H$ in (\ref{Hamiltonian}) has the same 
scaling dimension in $\vect{b}$ and $\vectg{\theta}$ (either 1 or 2), one 
finds that 
\begin{equation}
\frac{d}{d\tau} (\vect{b}\cdot\vectg{\theta})=\vectg{\theta}.\frac{\partial 
{\cal H}}{\partial \vectg{\theta}}-\vect{b}.\frac{\partial {\cal H}}
{\partial \vect{b}}=0\,,
\end{equation}
which shows that the overlap $\vect{b}\cdot\vectg{\theta}$ between the
physical and auxiliary fields is also conserved, together with $\cal H$,
along an extremal trajectory. This property reflects the scaling invariance
of the stochastic cascade processes we have in mind. We should at this 
point insist on the fact that, in contrast to instantons in the framework of
equilibrium statistical mechanics or quantum mechanics, equations for 
extremal trajectories in stochastic dynamical systems describe the real
motion of the physical field in a particular ``optimal'' realization of the
noise. The comparison of equations (\ref{Ito_tau}) and 
(\ref{extremalesMSR1}) shows indeed that the following relation holds
between $\vectg{\xi}$ and $\vectg{\theta}$~:
\begin{equation}
\vectg{\xi}=(\vect{b}\cdot\vect{b})^{1/2}\,^{t}\!B\vectg{\theta}\,.
\label{lienxitheta}
\end{equation}

Like their deterministic parent (\ref{GOYtau}), the equations of motion 
(\ref{extremalesMSR1}) and (\ref{extremalesMSR2}) sustain formally 
traveling wave-like solutions, such that
\begin{eqnarray}
\label{Floquet1}
b_{n+1}(\tau +T)&=&\exp AT\,b_{n}(\tau )\,,\\
\theta_{n+1}(\tau +T)&=&\exp -AT\,\theta_{n}(\tau )\,,
\label{Floquet2}
\end{eqnarray}
whose scaling exponent $z=\frac{AT}{\log Q}$ is expected to be now related
to the overlap $\mu_{1}=\vect{b}\cdot\vectg{\theta}$ (with $z=z_{0}$ for
$\mu_{1}=0$, in the absence of noise). However there is little hope to find
these solutions by a direct forward in time integration of 
(\ref{extremalesMSR1}) and (\ref{extremalesMSR2}), as could be done
 successfully for equation (\ref{GOYtau}). This is because the auxiliary field
 $\vectg{\theta}$ intrinsically propagates ``backward'' in time, as is clear
from the discretized version of (\ref{extremalesMSR2}) (deduced from 
the extremization of (\ref{MSRaction1})).
In the present problem, we have observed numerically 
that regular Floquet states emerge as dynamical attractors of 
(\ref{extremalesMSR1}) and (\ref{extremalesMSR2}) only for rather
high values of $\mu_{1}$ (otherwise the system evolves in a chaotic
manner).  They form a branch of solutions definitely distinct from the
one to be obtained in the next Section and correspond presumably to
local maxima of the action rather than the local minima of interest
to us. 

\section{An iterative method for computing self-similar instantons}
\label{ExtSol}
\subsection{Theory}
\label{ExtSoltheory}
The previous considerations suggest that equations (\ref{extremalesMSR1})
and (\ref{extremalesMSR2}) should not be treated on the same footing. 
The careful examination of physical properties that instantons should 
possess will give us keys for computing them.
Assume for a while that a solution has been found, obeying to 
(\ref{Floquet1}) and (\ref{Floquet2}).  We note $\vect{b}^{0}(\tau )$
and $\vectg{\xi}^{0}(\tau )$ the corresponding configurations of 
$\vect{b}$ and $\vectg{\xi}$. The linearization of the equation of motion
(\ref{extremalesMSR1}) at fixed noise leads to the following evolution of
fluctuations $\vectd{\delta}{ b}$ of $\vect{b}$ around $\vect{b}^{0}$~:
\begin{equation}
\label{linearisee1}
\frac{d}{d\tau }\vectd{\delta}{b}={\cal L}\vectd{\delta b}=
{\cal M}\vectd{\delta}{ b}+\left.(\vectd{\delta}{ b}\cdot\partial _{\vect{b}})
\left((\vect{b}\cdot\vect{b})^{1/2}\,B\vectg{\xi}^{0}\right)\right
\vert_{b^{0}}\,.
\end{equation}
The periodicity properties of the linear operator $\cal L$ ensure that
the fluctuations of $\vect{b}$ may be decomposed on a complete set
of eigendirections $\vectg{\Psi}_{ir}(\tau )$ 
 evolving according to (\ref{linearisee1}) and such that
\begin{equation}
\label{rightvect1}
\vectg{\Psi}_{ir}(\tau +T)={\e}^{\sigma_{i}T}\,{\cal T}_{+1}\,
\vectg{\Psi}_{ir}(\tau )\,,
\end{equation}
where ${\cal T}_{+1}$ denotes translation by one unit in the right direction
along the shell lattice. In practice we shall have to work with a finite
number of shells $d$ and, in order to get rid of boundary effects, it will be
necessary to fully periodize the shell lattice~: the index $i$ then runs between 1 and $d$ and the translation operator is easy to represent
as a matrix.  Formally, the $\vectg{\Psi}_{ir}$'s can be determined at
time $\tau =0$ by diagonalizing the Floquet operator~:
\begin{equation}
\label{rightop}
U_{T}={\cal T}_{-1}\,{\buildrel \leftarrow \over {\exp}}
\int _{0}^{T} {\cal L}\,d\tau\,,
\end{equation}
where $\buildrel \leftarrow \over {\exp}$ is a chronologically
 time ordered product (initial time on the right).
 One observes that $\vect{b}^{0}$ satisfies 
(\ref{rightvect1}) with a time averaged Lyapunov exponent $\sigma =A$.\\

 We claim now that every initial condition
$\vect{b}(0)$ evolving in the configuration of noise $\vectg{\xi}_{0}$
should be attracted towards the instantonic trajectory. If it were not true,
some perturbations would be able to grow in the comoving frame of the pulse,
thereby generating scaling exponents larger than $z$ at no cost in the action,
in contradiction with the hypothesis that the optimal realization of a 
singularity of exponent $z$ has been found. This strong criterion of
dynamic stability is another way of stating that the Cram\'er function
should be insensitive to the details of the production of pulses in the forcing
range. It implies that $A$ is an upper bound for the real part of the
 $\sigma_{i}$'s. Arranging the 
eigendirections $\vectg{\Psi}_{ir}$ in order of decreasing $\rm{Re}\, 
\sigma _{i}$, we are therefore led to identify $\vect{b}^{0}(\tau )$
with $\vectg{\Psi}_{1\,r}(\tau )$. In the case of zero noise where we recover
the deterministic solution of Section \ref{def} (with $\vectg{\xi}^{0}=
\vect{0}$ in both (\ref{extremalesMSR1}) and (\ref{linearisee1})), the time
derivative $\frac{d\vect{b}^{0}}{d\tau }$ is also solution of 
(\ref{linearisee1}) with the same Lyapunov exponent as $\vect{b}^{0}$,
$\sigma =A$ ($=A_{0}$ in this case). In this limit we would naturally
define $\vectg{\Psi}_{2r}(\tau )$ as $\frac{d\vect{b}^{0}}{d\tau }$. This
property is lost in the more general situation of a non vanishing noise,
because $\vectg{\xi}^{0}$ is not time-invariant. What remains true 
however is the fact that $\vect{b}^{0}$ and $\frac{d\vect{b}^{0}}{d\tau }$ 
still span the set of ``coherent'' fluctuations which do not affect the shape
of the pulse but modify its height and position. \\

By turning now our attention to the linear dynamics dual to
(\ref{linearisee1}) we shall come close to the equation 
(\ref{extremalesMSR2}). Let us indeed consider the equation of motion
\begin{equation}
\label{dual1}
\frac{d\vectg{\theta}}{d\tau }=-^{t}\!{\cal L}\,\vectg{\theta }\,,
\end{equation}
where in order to limit the proliferation of symbols, we keep the same
notation $\vectg{\theta }$ for the new auxiliary field, although it is only
in particular circumstances, to be clarified below, related to the 
$\vectg{\theta }$ of equations (\ref{extremalesMSR1}) and 
(\ref{extremalesMSR2}). From (\ref{linearisee1}) we get~:
\begin{equation}
\frac{d\vectg{\theta}}{d\tau }=-^{t}\!{\cal M}\,\vectg{\theta}
-\left.\partial _{\vect{b}}\left((\vect{b}\cdot\vect{b})^{1/2}\,\vectg{\theta}.
B\vectg{\xi}^{0}\right)\right\vert_{\vect{b}^{0}}\,.
\end{equation}
The dual dynamics enables one to construct a basis of left eigenvectors
$\vectg{\Psi}_{il}(\tau )$ (with $1\leq i\leq d$) satisfying
\begin{equation}
\label{ortho}
\vectg{\Psi}_{il}(\tau +T)={\e}^{-\sigma_{i}T}\,{\cal T}_{+1}\,
\vectg{\Psi}_{il}(\tau )\,,
\end{equation}
as well as the following orthogonality conditions with the members of the
first basis~:
\begin{equation}
\vectg{\Psi}_{il}(\tau )\cdot\vectg{\Psi}_{jr}(\tau )=\delta_{ij}\,,
\end{equation}
at every time. The vectors $\vectg{\Psi}_{il}(0)$ are determined by diagonalizing the adjoint Floquet operator 
\begin{equation}
\label{leftop}
^{t}U_{T}={\buildrel \rightarrow \over {\exp}}
\int _{0}^{T} {^{t}\!{\cal L}}\,d\tau\;{\cal T}_{+1}\,,
\end{equation}
and enforcing the normalization condition (\ref{ortho}) at time $\tau =0$ 
($\buildrel \rightarrow \over {\exp}$ is now an anti-chronologically time
ordered product). We may note at this point that the first left eigenvector
$\vectg{\Psi}_{1l}$ is, in the generic case of non-zero noise, the only one to
display the scaling behaviour anticipated for $\vectg{\theta}^{0}$ 
according to (\ref{Floquet2}), since its Lyapunov exponent equals 
$-\sigma_{1}=-A$. We conclude that the auxiliary equation 
(\ref{extremalesMSR2}) in the restricted manifold of self-similar
solutions is tantamount to the relation~:
\begin{equation}
\label{selfcons1}
\vectg{\theta}^{0}(\tau )=\mu_{1}\,\vectg{\Psi}_{1l}(\tau )\,,
\end{equation}
where the multiplicative constant $\mu_{1}$ is nothing but the
overlap $\vectg{\theta}^{0}\cdot\vect{b}^{0}$ ($=\mu_{1}\vectg{\Psi}_{1l}
(\tau )\cdot\vectg{\Psi}_{1r}(\tau )=\mu_{1}$), which was shown before
to be indeed a conserved quantity. This claim is further confirmed by
rewriting the original equation (\ref{extremalesMSR2}) as
\begin{eqnarray*}
\frac{d\vectg{\theta}}{d\tau }&=&-^{t}\!{\cal M}\,\vectg{\theta}
-(\vectg{\theta}\cdot B\,^{t}\!B\vectg{\theta})\vect{b}
-(\vect{b}\cdot\vect{b})\,\vectg{\theta}.\frac{1}{2}[(\partial_{\vect{b}}\,B)\,
^{t}\!B+B\,(\partial_{\vect{b}}\,^{t}\!B)]\vectg{\theta}\\
 &=&-^{t}\!{\cal M}\,\vectg{\theta}
-(\vectg{\theta}\cdot B\,^{t}\!B\vectg{\theta})\vect{b}-(\vect{b}\cdot\vect{b})
\,\vectg{\theta}.(\partial_{\vect{b}}\,B)\,^{t}\!B\vectg{\theta}\,.
\end{eqnarray*}
Putting back 0 superscripts and reintroducing $\vectg{\xi}^{0}$ by using  (\ref{lienxitheta}), we arrive at 
\begin{equation}
\frac{d\vectg{\theta}^{0}}{d\tau }=-^{t}\!{\cal M}\,\vectg{\theta}^{0}
-(\vectg{\theta}^{0}\cdot B\vectg{\xi}^{0})\frac{\vect{b}^{0}}
{(\vect{b}^{0}\cdot\vect{b}^{0})^{1/2}}-(\vect{b}^{0}\cdot\vect{b}^{0})^{1/2}
\,\partial_{\vect{b}}(\vectg{\theta}^{0}\cdot B\vectg{\xi}^{0})\,,
\end{equation}
which shows that $\vectg{\theta}^{0}$ obeys the dual dynamics defined
by equation (\ref{dual1}).\\

Having interpreted (\ref{extremalesMSR2}) as a condition of 
self-consistency for the conjugate momentum $\vectg{\theta}$ expressed
by (\ref{selfcons1}), we could contemplate the following Newton-like
procedure for catching numerically self-similar instantons. First make a
guess for $\vectg{\xi}$ in the form of a traveling wave ($\vectg{\xi}^{in}
(\tau +T)={\cal T}_{+1}\,\vectg{\xi}^{in}(\tau )$), integrate 
(\ref{extremalesMSR2}) forward in time in order to determine the
asymptotic traveling state reached by $\vect{b}$ in the prescribed
configuration of the noise. Then compute $\vectg{\Psi}_{1l}$ from the
diagonalization of $^{t}U_{T}$ (or from running (\ref{dual1}) backward in time in order to let emerge the eigendirection of lowest growth rate), employ (\ref{selfcons1}) for producing a new configuration of 
$\vectg{\theta}$ (and thereby $\vectg{\xi}$) and iterate this loop many 
times at a fixed value of the overlap $\mu_{1}$ until convergence is
achieved. However two major difficulties call for an improvement of the
method~: they both have to do with the stability of the trajectory upon
time reparametrization. First we do not know the speed (or the inverse
period $T^{-1}$) of the final traveling wave which must carry together
$\vect{b}$ and $\vectg{\xi}$. Therefore, when performing the first step
of the iterative loop, we must allow continuous time reparametrization of 
our Ansatz for the noise in order to fine tune the speeds of the two pulses
formed by $\vect{b}$ and $\vectg{\xi}$ and let both terms in the right 
hand side of (\ref{extremalesMSR1}) be always relevant. It will be 
explained in the next Subsection how this goal can be achieved in practice.
The second difficulty is much more serious than the preceding one and in 
the way of getting around it resides perhaps the most tricky part of this
work. 
The point is that traveling wave solutions to (\ref{extremalesMSR1})
and (\ref{extremalesMSR2}) may perfectly have a non zero pseudo-energy
$\cal H$, while we are looking for the particular ones with ${\cal H}=0$. 
We shall be able to fulfill asymptotically the two conditions 
$\vect{b}.
\vectg{\theta}=\mu_{1}$ and ${\cal H}=0$, if and only if our iterative
guess for $\vectg{\theta}$ is constructed within a two-dimensional
space rather than a unidimensional one as in the naive proposal
made above. 
 For this purpose, we are going to embed the linearized
dynamics (\ref{linearisee1}) into a new one which admits the time
translation mode $\frac{d\vect{b}^{0}}{d\tau}$ as a true eigenstate, of the
same growth factor $A$ as $\vect{b}^{0}$, restoring thereby the
symmetries present in the absence of noise. We shall do that in the most
economical way, from both formal and numerical points of view, by
substituting $\vectg{\Phi}_{2r}=\frac{d\vect{b}^{0}}{d\tau}$ to
$\vectg{\Psi}_{2r}$, i.e., the eigendirection along which the fluctuations
of $\vect{b}$ around $\vect{b}^{0}$ are the less stable. \\

The left eigenvector $\vectg{\Psi}_{2l}$ is first rescaled as
$\vectg{\Phi}_{2l}=\frac{\vectg{\Psi}_{2l}}{\vectg{\Psi}_{2l}.
\vectg{\Phi}_{2r}}$ (which makes sense as long as $\vectg{\Psi}_{2l}.
\vectg{\Phi}_{2r} \neq 0$, a condition always found to be satisfied in
practice), so that $\vectg{\Phi}_{2l}$ has a unit overlap with
$\vectg{\Phi}_{2r}$, while being orthogonal to all other right
eigenvectors $\vectg{\Psi}_{ir}$ with $i\neq 2$ (which will be noted
$\vectg{\Phi}_{ir}$ from now on). One then considers the modified
linear dynamics~:
\begin{equation}
\label{linearisee2}
\frac{d\vectd{\delta}{b}}{d\tau }=\tilde{\cal L} \vectd{\delta}{b}
={\cal L} \vectd{\delta}{ b}+(\vect{b}^{0}\cdot\vect{b}^{0})^{1/2}\,
(\vectg{\Phi}_{2l}\cdot\vectd{\delta}{ b})\,B\,\frac{d\vectg{\xi}^{0}}{d\tau }\,.
\end{equation}
It is easily checked that $\vectg{\Phi}_{2r}=\frac{d\vect{b}^{0}}{d\tau}$
obeys (\ref{linearisee2}), since equation (\ref{extremalesMSR1}) yields
upon time derivation~:
\begin{equation}
\frac{d\vectg{\Phi}_{2r}}{d\tau }
={\cal L} \vectg{\Phi}_{2r}+(\vect{b}^{0}\cdot\vect{b}^{0})^{1/2}\,
B\,\frac{d\vectg{\xi}^{0}}{d\tau }=\tilde{\cal L} \vectg{\Phi}_{2r}\,.
\end{equation}
It is also trivially seen that the other vectors $\vectg{\Phi}_{ir}=
\vectg{\Psi}_{ir}$ for $i\neq 2$ keep the same evolution under 
(\ref{linearisee2}) as under (\ref{linearisee1}). The dual dynamics reads
now~:
\begin{equation}
\label{dual2}
\frac{d\vectg{\theta}}{d\tau }=-^{t}\!\tilde{\cal L} \vectg{\theta}
=-^{t}\!{\cal L} \vectg{\theta }-(\vect{b}^{0}\cdot\vect{b}^{0})^{1/2}\,
(\vectg{\theta}\cdot B\,\frac{d\vectg{\xi}^{0}}{d\tau })\vectg{\Phi}_{2l}\,.
\end{equation}
It leads to a new family of left eigenvectors $\vectg{\Phi}_{il}$, dual to
the direct basis, whose second member  $\vectg{\Phi}_{2l}$ has been 
defined above and the others relate to their original counterparts
$\vectg{\Psi}_{il}$ as 
\begin{equation}
\label{lienPhiPsi}
\vectg{\Phi}_{il}=\vectg{\Psi}_{il}-(\vectg{\Psi}_{il}\cdot \vectg{\Phi}_{2r})
\vectg{\Phi}_{2l}\,.
\end{equation}

Although the $\vectg{\Phi}_{il}$'s were introduced as a rather formal
trick, it should be emphasized that $\vectg{\Phi}_{1l}$ and 
$\vectg{\Phi}_{2l}$ have an appealing physical meaning. Parameterizing
a perturbed trajectory for $\vect{b}$ as $\vect{b}={\e}^{\delta {\rm ln}
b(\tau )}\,\vect{b}^{0}(\tau +\delta \tau (\tau ))+\vectd{\delta}{b}_{inc}$,
where the ``incoherent'' part of fluctuations $\vectd{\delta}{b}_{inc}$ is
bound to be a linear superposition of the less dangerous modes 
$\vectg{\Phi}_{ir}$ for $i\geq 3$, one has indeed, to linear order in
$\vectd{\delta}{ b}$,
\begin{eqnarray}
\label{growthpert}
\delta \mbox{ln}b &=&\vectg{\Phi}_{1l}\cdot\vectd{\delta}{ b}\,,\\
\delta \tau &=&\vectg{\Phi}_{2l}\cdot\vectd{\delta}{ b}\,.
\label{speedpert}
\end{eqnarray}
These two relations will be useful in the computation of quadratic
fluctuations to be presented in Section \ref{fluct}. They show that
by projecting out the multidimensional fluctuation field 
$\vectd{\delta}{ b}$
onto the two vectors $\vectg{\Phi}_{1l}$ and $\vectg{\Phi}_{2l}$, one has 
access to the most relevant part of it affecting respectively the amplitude
and the time delay of the pulse constituting the instanton. In terms of
$\vectg{\Phi}_{1l}$ and $\vectg{\Phi}_{2l}$, the self-consistency
condition (\ref{selfcons1}) for $\vectg{\theta}^{0}$ together with the
requirement of zero pseudo-energy $\cal H$ take the following form~:
\begin{equation}
\label{selfcons2}
\vectg{\theta}^{0}(\tau )=\mu_{1}\vectg{\Phi}_{1l}(\tau )
+\mu_{2}(\tau )\vectg{\Phi}_{2l}(\tau )\,,
\end{equation}
where
\begin{equation}
\label{def_mu2}
\mu_{2}(\tau )=\frac{1}{2}\vectg{\theta}^{0}\cdot B\,^{t}\!B\,
\vectg{\theta}^{0}=\frac{1}{2}\vectg{\xi}^{0}\cdot\vectg{\xi}^{0}\,.
\end{equation}
Since from (\ref{selfcons2}), $\mu_{2}(\tau )=\vectg{\Phi}_{2r}.
\vectg{\theta}^{0}$, and from the equation of motion 
(\ref{extremalesMSR2}), $\vectg{\Phi}_{2r}\cdot\vectg{\theta}^{0}=
{\cal H}+\frac{1}{2}\vectg{\xi}^{0}\cdot\vectg{\xi}^{0}$, the relation 
(\ref{def_mu2}) is just a way of restating ${\cal H}=0$. That 
(\ref{selfcons1}) implies (\ref{selfcons2}) results from the general link 
between $\vectg{\Psi}_{1l}$ and $\vectg{\Phi}_{1l}$ 
(see (\ref{lienPhiPsi})).
The reverse is true only under the supplementary condition of constant
$\cal H$ or 
$\mu_{2}(\tau )=C^{te}+\frac{1}{2}\vectg{\xi}^{0}\cdot\vectg{\xi}^{0}$, which
is guaranteed by (\ref{def_mu2}). It is proven by checking that in that case
$\vectg{\theta}^{0}(\tau )$, as given by (\ref{selfcons2}), obeys, as it
should, (\ref{dual1})~:
\begin{eqnarray*}
\frac{d\vectg{\theta}^{0}}{d\tau }&=& -^{t}\!\tilde{\cal L}\vectg{\theta}^{0}
+\frac{d\mu_{2}}{d\tau }\,\vectg{\Phi}_{2l}\\
&=&-^{t}\!{\cal L}\vectg{\theta}^{0} -(\vect{b}^{0}\cdot\vect{b}^{0})^{1/2}
(\vectg{\theta}^{0}\cdot B\,\frac{d\vectg{\xi}^{0}}{d\tau })\vectg{\Phi}_{2l}
+\frac{d\mu_{2}}{d\tau }\,\vectg{\Phi}_{2l}\\
&=&-^{t}\!{\cal L}\vectg{\theta}^{0} +\frac{d}{d\tau }(\mu_{2}-
\frac{1}{2}\vectg{\xi}^{0}\cdot\vectg{\xi}^{0})\,\vectg{\Phi}_{2l}
=-^{t}\!{\cal L}\vectg{\theta}^{0}+\frac{d{\cal H}}{d\tau }\,\vectg{\Phi}_{2l}\,.
\end{eqnarray*}

The great advantage of (\ref{selfcons2}) and (\ref{def_mu2}) with respect 
to (\ref{selfcons1}) is that this couple of equations lends itself to 
iterative procedures leading inexorably to a fixed point of zero 
pseudo-energy, a task seemingly out of reach before. There is some unavoidable
arbitrariness in the construction proposed here, concerning in particular
 the definition of the vector $\vectg{\Phi}_{2l}$, about which the
reader may feel a little uncomfortable. We suspect that these unwanted 
features do not affect the final results since the original equations to be
solved as well as the corresponding conserved quantities all have a clear 
mathematical definition where $\vectg{\Phi}_{1l}$ and $\vectg{\Phi}_{2l}$
merge together into $\vectg{\Psi}_{1l}$. 

\subsection{Practical implementing and results}
\label{ExtSolimplementing}

\begin{figure}[l]
\centerline{\psfig{file=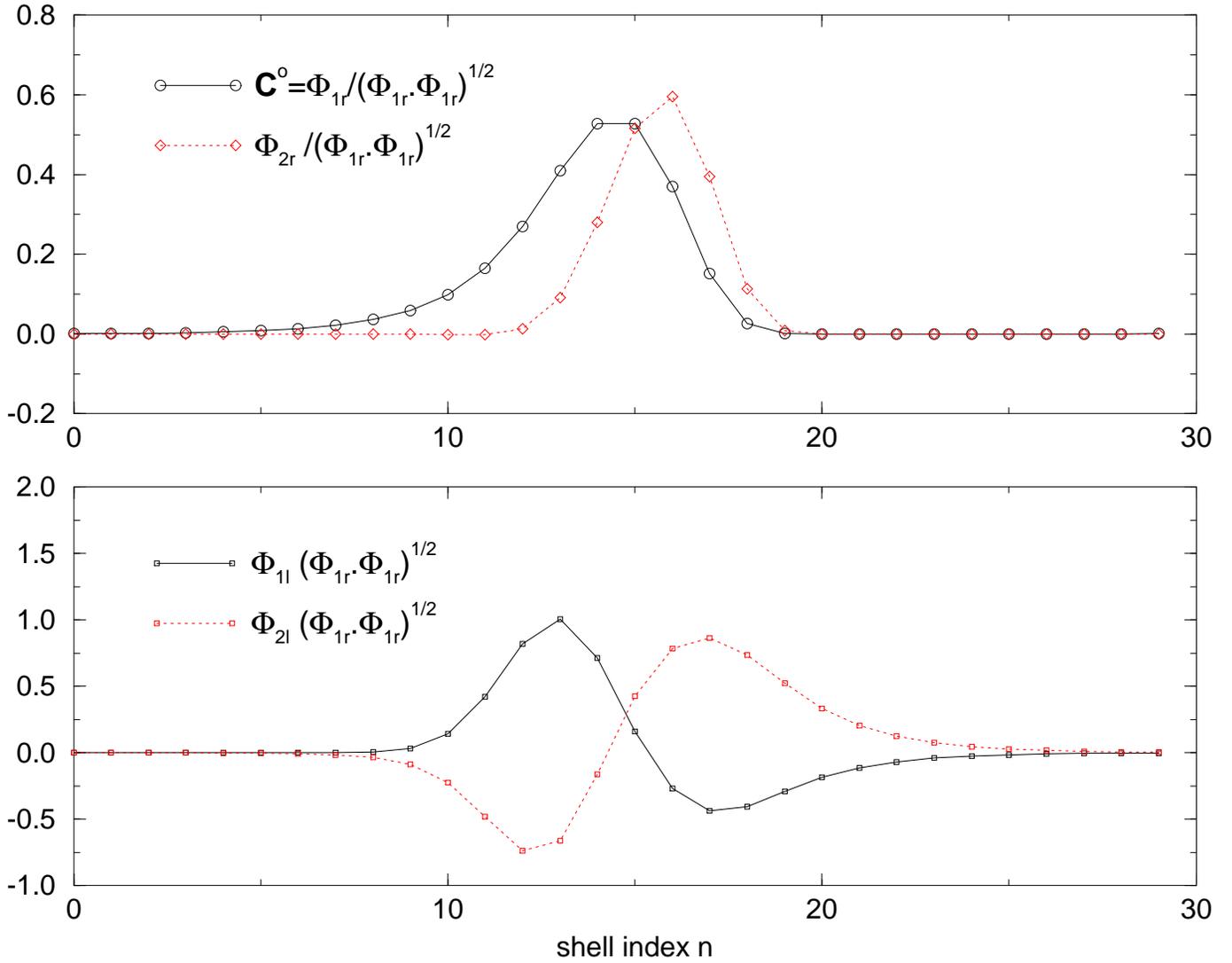}}
\caption{On the upper picture (resp. lower) we plot the configurations at a 
given instant of the right (resp. left) eigenmodes in the 
subspace of maximum Lyapunov exponent $A_0$ around the deterministic solution.}
\label{Fig.1}
\end{figure}

The action density $s_{0}(z)$ could be computed successfully for the three
stochastic models defined in Section \ref{def} using the iterative scheme
outlined before. The shell lattice was first mapped onto a circle of
$d$ sites, with $d$ typically ranging between 20 and 30. Finite size
effects turn out to be completely negligible at such lengths of the chain,
due to the strongly localized structure of the instantons. To start the
computation, we make a guess for both the unit field $\vect{C}(\tau )=
\frac{\vect{b}(\tau )}{(\vect{b}\cdot\vect{b})^{1/2}(\tau )}$ and the noise
$\vectg{\xi }(\tau )$ called henceforth $\vect{C}^{in}(\tau )$ and 
$\vectg{\xi }^{in}(\tau )$. They are such that
\begin{equation}
\label{periodicity1}
C^{in}_{n+1}(\tau +T^{in})=C^{in}_{n}(\tau ), \quad \quad
\xi^{in}_{n+1}(\tau +T^{in})=\xi^{in}_{n}(\tau ), 
\end{equation}
and
\begin{equation}
C^{in}_{n+d}(\tau )=C^{in}_{n}(\tau ), \quad \quad
\xi^{in}_{n+d}(\tau )=\xi^{in}_{n}(\tau ).
\end{equation}
Furthermore, the noise is normalized in such a way that the overlap
$\vect{b}\cdot\vectg{\theta}=\vect{C}\cdot B^{-1}[\vect{C}]\vectg{\xi}$, takes 
on a prescribed value $\mu_1 $ held as a control parameter during all the 
steps of the computation. A possible and convenient choice would be for
instance $\vect{C}^{in}(\tau )=\vect{C}^{0}(\tau )$, where $\vect{C}^{0}
(\tau )$ is the deterministic solution of scaling exponent $z_{0}$, and
$\vectg{\xi}^{in}(\tau )=\mu_1 (\vect{b}^{0}\cdot\vect{b}^{0})^{1/2}
B[\vect{C}^{0}] \vectg{\Phi}^{0}_{1l}(\tau )$, where 
$\vectg{\Phi}^{0}_{1l}(\tau )$ is the left eigenvector dual to 
$\vect{b}^{0}(\tau )$, i.e., in the absence of noise. Fig. \ref{Fig.1}
shows how both vectors $\vectg{\Phi}^{0}_{1l}$ and 
$\vectg{\Phi}^{0}_{2l}$ look like at a given time. Their shapes will
in fact little evolve as we let the scaling exponent $z$ depart
from $z_{0}$. \\
In order to allow time reparametrization of the trajectory, we first
get an estimate of the instantaneous position of the
pulse along the shell axis in our trial configuration by computing
the following quantity
\begin{equation}
\label{def_centermass}
n^{in}(\tau )=\sum_{n=0}^{d-1}n[d]\left(C^{in}_{n}(\tau )\right)^{2}.
\end{equation}
The notation $n[d]$ recalls that, due to cyclic boundary conditions, the 
shell index $n$ is now only defined modulo $d$ and that in practice a
continuous determination of this integer should be adopted close to the
center of the pulse which contributes mostly to 
the right hand side of (\ref{def_centermass}). One has by construction
$n^{in}(\tau +T^{in})=n^{in}(\tau )+1 \pmod{d} $. Having recorded
$n^{in}(\tau )$ and $\vectg{\xi}^{in}(\tau )$ during a whole period 
$T^{in}$, we integrate forward in time the nonlinear 
evolution equation for $\vect{C}$ deduced from (\ref{Ito_tau}) by projecting
out the longitudinal part of its right hand side~:
\begin{equation}
\label{dynC}
\frac{d\vect{C}}{d\tau }=\left\{\vect{N}[\vect{C}](\tau )+B[\vect{C}]
(\tau )\vectg{\xi}^{in}(\tau ')\right\}_{\perp},
\end{equation}
Note that the subindex $\Gamma $ disappeared from the nonlinear kernel
$\vect{N}_{\Gamma }$ because the Ito-drift term being linear in $\Gamma $
does not matter in the computation of the action to leading order (we shall
see in the next Section how to handle it to next to leading order). The
most salient feature of (\ref{dynC}) is that the noise configuration is
evaluated in relation not to the time $\tau $ but rather to the actual 
instantaneous position of the pulse. This means that the time $\tau '(\tau )$ 
is automatically delayed or advanced with respect to $\tau $, according to
the recipe 
\begin{equation}
\label{lientemps}
n^{in}(\tau ')=n(\tau ) \quad \mbox{or}\quad \tau '=(n^{in})^{-1}
[n(\tau )].
\end{equation}
After integrating (\ref{dynC}) long enough, a new traveling wave state 
$\vect{C}^{out}(\tau )$, $\vectg{\xi}^{out}(\tau )=\vectg{\xi}^{in}
(\tau '(\tau ))$ will usually
emerge, of period $T^{out}$ possibly different from $T^{in}$ and 
averaged growth factor
\begin{equation}
A^{out}=\frac{1}{T^{out}} \int_{\tau}^{\tau +T^{out}}
\left(\vect{N}[\vect{C}^{out}]+B[\vect{C}^{out}]\vectg{\xi}^{out}\right).
\vect{C}^{out}\,d\tau .
\end{equation}
The vectors $\vectg{\Psi}_{1l}(\tau )$, $\vectg{\Psi}_{2l}(\tau )$ 
are then identified as the two eigenvectors of $^{t}U_{T^{out}}$ (defined in (\ref{leftop})) of smallest (real negative) Lyapunov coefficient and the
corresponding $\vectg{\Phi}_{1l}(\tau )$, $\vectg{\Phi}_{2l}(\tau )$
constructed as linear combinations of them obeying for all times the
following relations
\begin{eqnarray*}
\vectg{\Phi}_{1l}\cdot\vect{b}^{out}&=&\vectg{\Phi}_{2l}.\frac{d\vect{b}^{out}}
{d\tau }=1 \\
\vectg{\Phi}_{2l}\cdot\vect{b}^{out}&=&
\vectg{\Phi}_{1l}.\frac{d\vect{b}^{out}}{d\tau }=0
\end{eqnarray*}
Finally, the trial noise configuration is renewed as $\vectg{\xi}^{in}(\tau )=
(\vect{b}^{out}\cdot\vect{b}^{out})^{1/2}\,
 {^{t}\!B}[\vect{C}^{out}]\vectg{\theta}^{in}$
with
\begin{equation} 
\vectg{\theta}^{in}(\tau )=\mu_1 \vectg{\Phi}_{1l}(\tau )+\mu_2 (\tau )
\vectg{\Phi}_{2l}(\tau ),
\end{equation}
where $\mu_2 (\tau )$ is determined upon imposing the condition of zero 
pseudo-energy on the trial solution ($\vectg{\theta}^{in}(\tau)$,
$\vect{b}^{out}(\tau)$)
\begin{equation}
\vectg{\theta}^{in}\cdot\vect{N}[\vect{C}^{out}]+\frac{1}{2}(\vect{b}^{out}.
\vect{b}^{out})\vectg{\theta}^{in}\cdot B[\vect{C}^{out}]\,{^{t}\!B}[\vect{C}^{out}]
\vectg{\theta}^{in}=0
\end{equation}
After setting $\vect{b}^{in}(\tau )=\vect{b}^{out}(\tau )$, we are ready
to repeat the operations described above as many times as needed until a fixed
point of the transformation (such that $\vect{b}^{out}(\tau )=\vect{b}^{in}
(\tau )$ and $\vectg{\xi}^{in}(\tau )=\vectg{\xi}^{out}(\tau )$) is reached, 
which solves the problem. The good stability properties of the algorithm,
as well as its iterative character, authorize a rather unsophisticated
handling of issues raised by the time discretization. As required by the Ito-
convention, the equation of motion (\ref{dynC}) was integrated using a 
first-order Euler scheme with a time step $\Delta \tau =\frac{T^{in}}{N}$ about
350 times smaller than the period. The time $\tau '$ was approximated as the
multiple of $\Delta \tau $ making the relation (\ref{lientemps}) best
satisfied. Similarly no higher order interpolation scheme was devised for
estimating with accuracy the output period $T^{out}$~: it was again simply
approximated as the multiple of $\Delta \tau $ making periodicity
conditions (\ref{periodicity1}) best satisfied for the output. However the 
time step $\Delta \tau $ was changed at each iteration of the loop so to
maintain the time resolution $N$ constant. The efficiency of the method was
 greatly improved, when seeking solutions of exponents $z$ far from $z_0$,
by increasing $z$ gradually (through an increase of the control parameter
$\mu_1 $) using as first guess solutions of lower but close scaling
exponent $z'$. In this way convergence toward satisfactory self-similar
solutions (of exponent $z$ varying by less than $10^{-5}$ under iteration
and pseudo-energy ${\cal H} \le 10^{-5}$) was attained in no more than 20
iterations.\\

\begin{figure}[l]
\centerline{\psfig{file=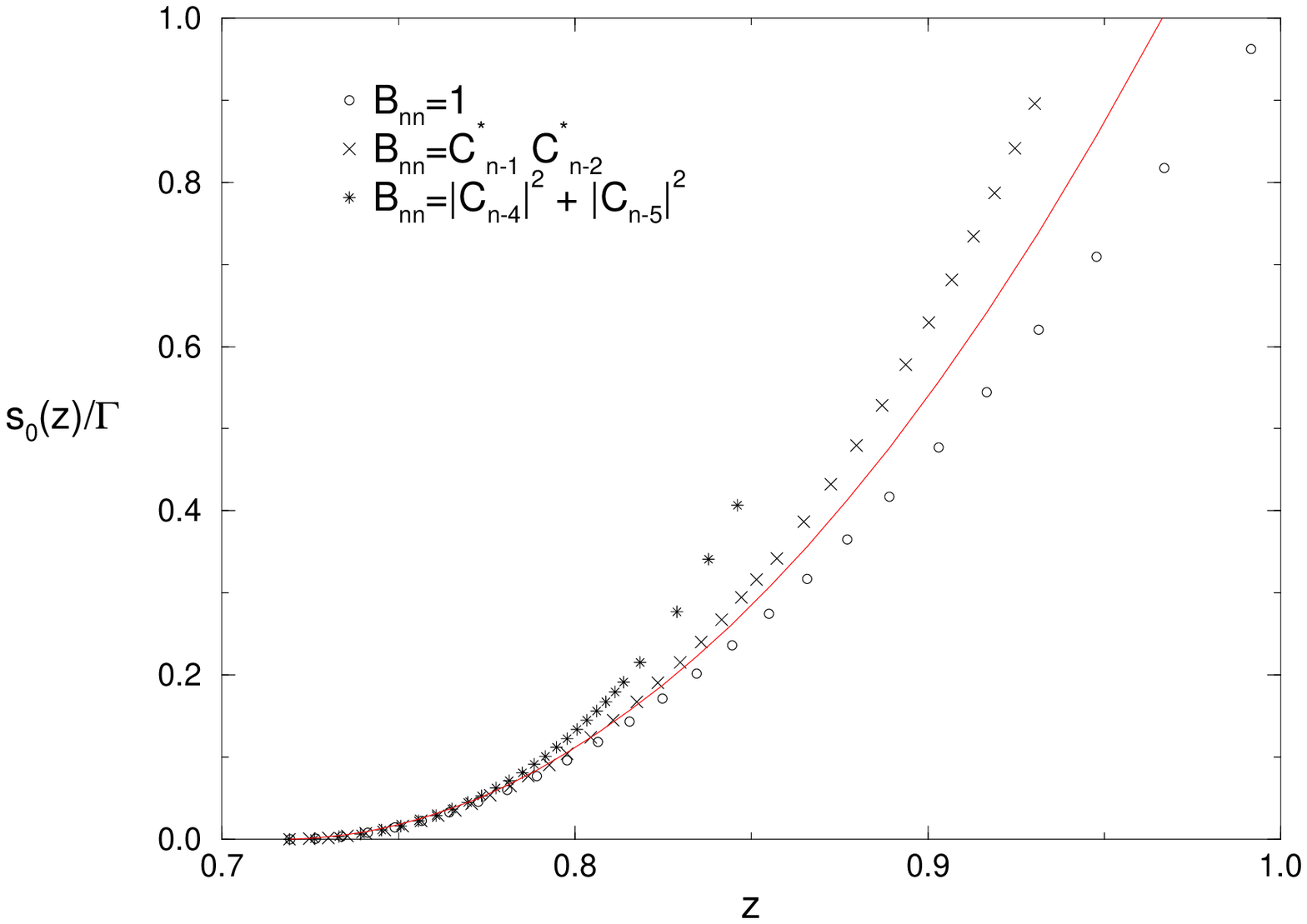}}
\caption{Evolution of the normalized action per unit cascade step  $s_0(z)/\Gamma$
, as a function of the effective scaling exponent $z$, for the three models 
studied in this paper.
As a guide for the eyes we show the  parabola (solid line) "tangent" to the 
curves at the deterministic minimum $z_0=0.72$. }
\label{Fig.2}
\end{figure}

\begin{figure}[l]
\centerline{\psfig{file=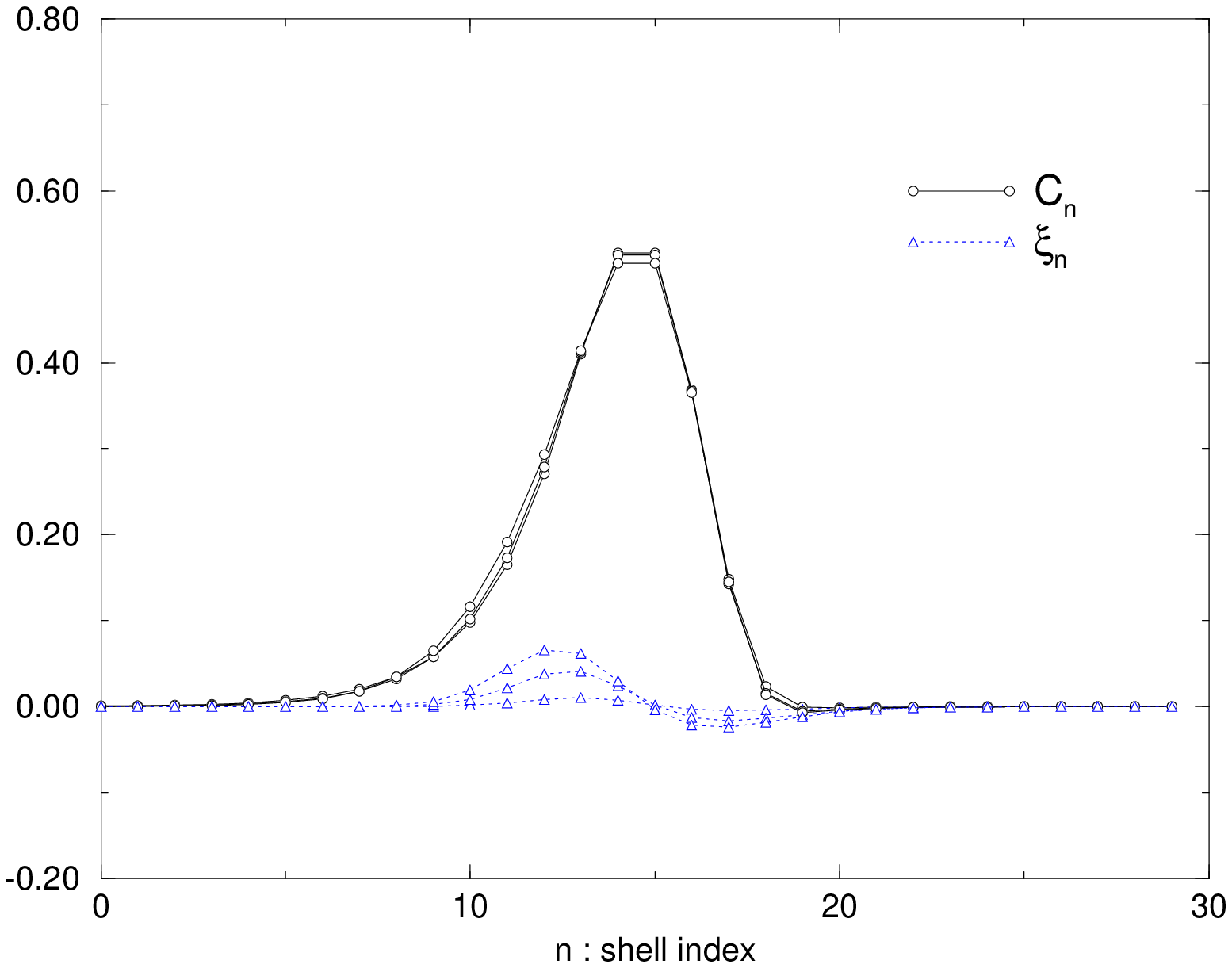}}
\caption{Configurations of the normalized coherent field $\vect{C}$ and the 
random force $B\vect{\xi}$ for three instantons of exponents equal to
$z=0.75,0.85,0.95$, obtained with the model (i) (according to the 
nomenclature defined in the text). Note that the $\vect{C}$ field is
only slightly deformed as $z$ increases. }
\label{Fig.3}
\end{figure}

\begin{figure}[l]
\centerline{\psfig{file=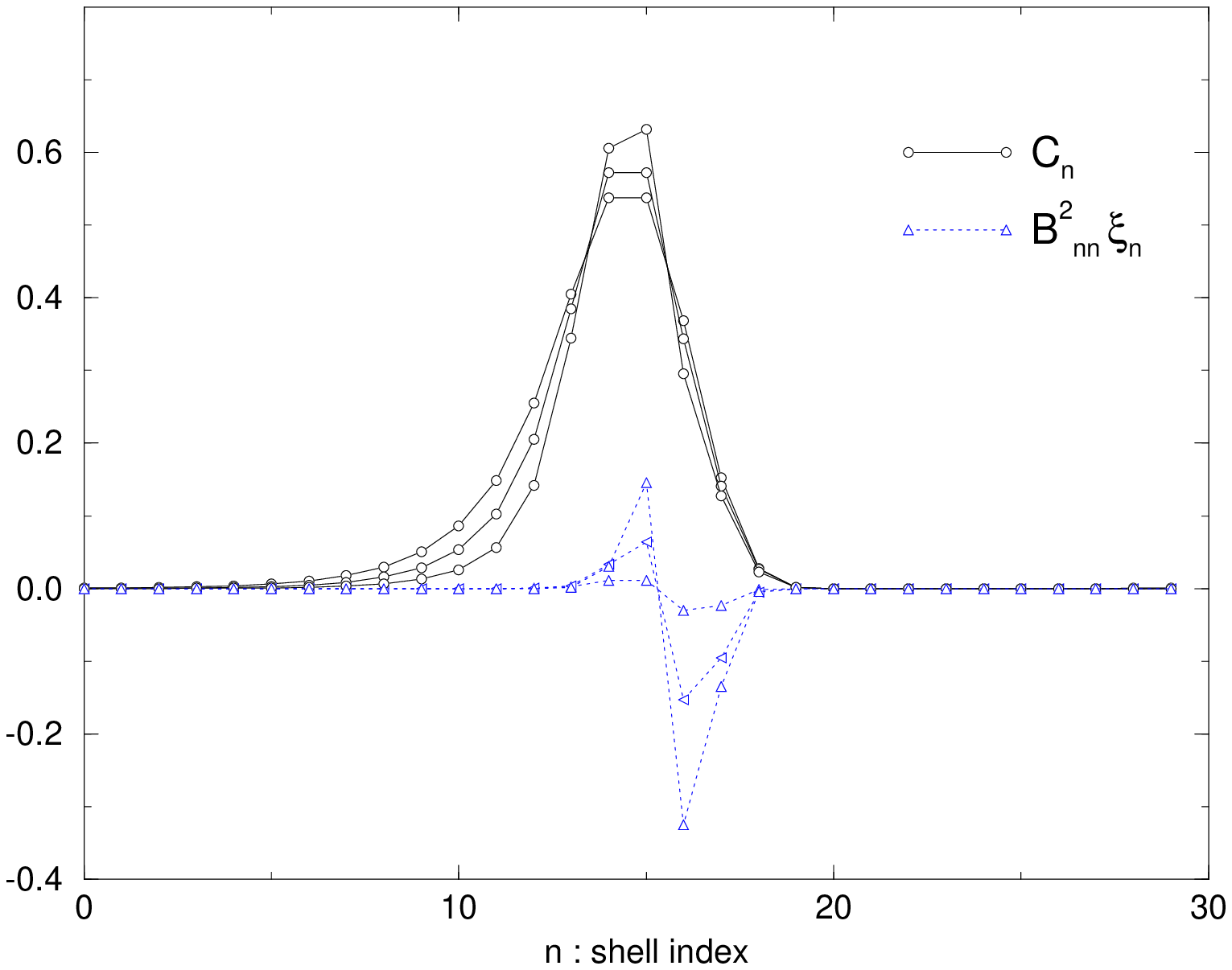}}
\caption{Same as in Fig.\protect{\ref{Fig.3}} but for the model (ii). Note 
the more pronounced deformation of $\vect{C}$ upon increasing $z$, with respect
to the previous case.  }
\label{Fig.4}
\end{figure}

\begin{figure}[l]
\centerline{\psfig{file=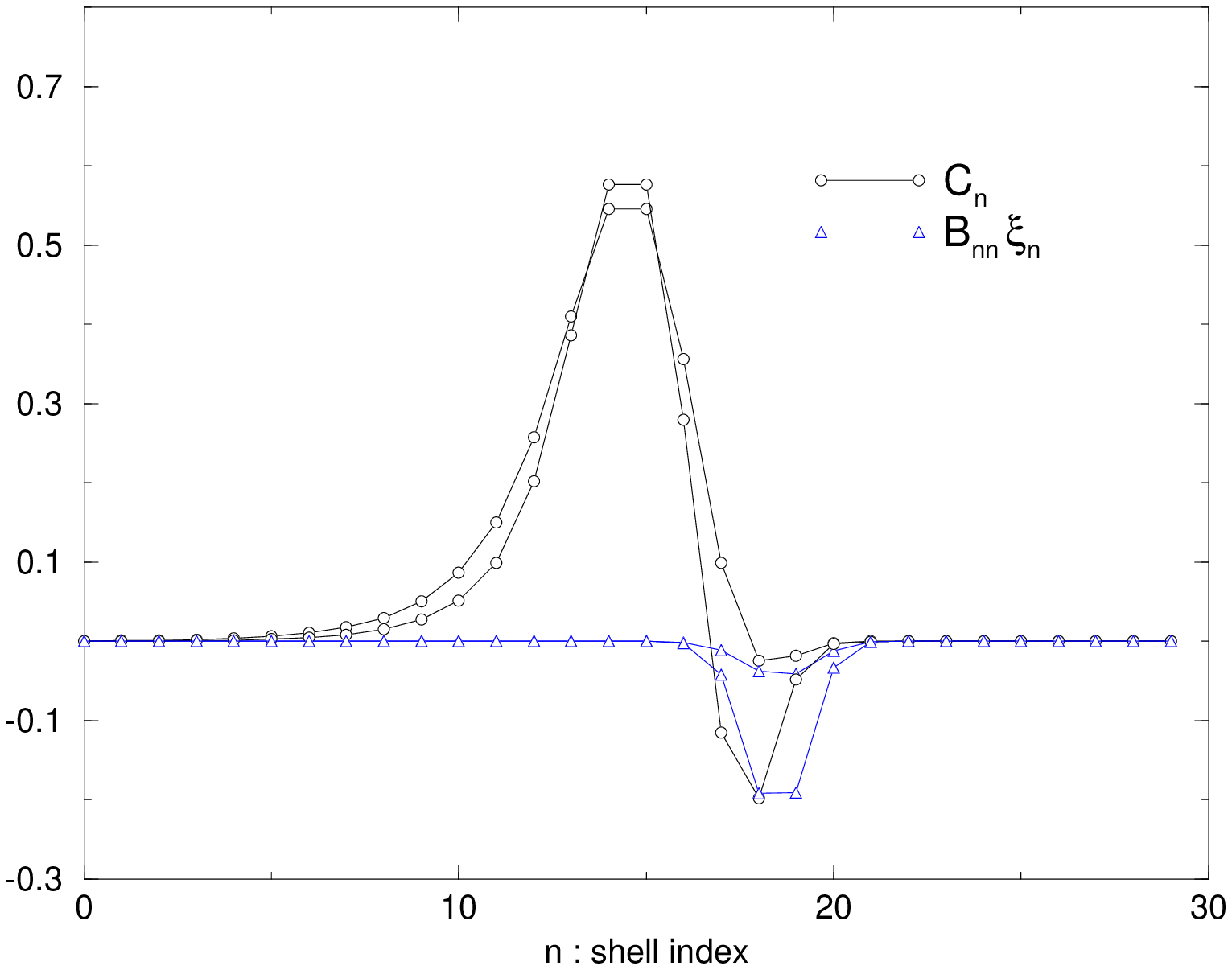}}
\caption{Same as in Fig.\protect{\ref{Fig.3}} but for the model (iii) and 
only instantons of exponent $z=0.75, 0.85$. The 
physical field now changes its sign at the leading edge of the pulse. }
\label{Fig.5}
\end{figure}

We turn now to the presentation of our results. Fig.~\ref{Fig.2} shows the
 action
density $s_{0}(z)$ for the three models (i), (ii) and (iii) defined in Section
\ref{def}. Values of $\Gamma $ were adjusted in order to provide the same
curvature of $s_{0}(z)/\Gamma $ at the bottom of the curves, reached
evidently at $z=z_{0}$. We see that the variations of the zeroth-order
action get sharper on the $z>z_0$ side (the only one displayed in 
Fig. \ref{Fig.2}), as one goes from model (i) to model (iii). Figures \ref{Fig.3},
\ref{Fig.4} and \ref{Fig.5}, referring respectively to models (i), (ii) and (iii), show
the normalized coherent field $\vect{C}$ and the random force $B\vectg{\xi}$ at increasing values of $z$ (0.75, 0.85 and 0.95). In all cases
the random force is found to be negative at the leading edge of the pulse, in
 agreement with the physical picture advocated in \cite{GD97}~: growth
can be enhanced only by frustrating the energy transfer processes. For 
model (iii), the coherent field itself gets negative at the forefront~: noise
in that case just helps to prepare the system in an initial condition 
consisting of a pulse and a negative well in front of it, which then collide. 
An interesting upshot of our computations is that models like (ii) or (iii)
involving only a local coupling between $\vect{b}$ and $\vectg{\xi}$ escape the disaster met in the framework of model (i), namely a cross-over
toward an asymptotic  linear growth of $s_{0}(z)$ with $z$, already 
perceptible in Fig. \ref{Fig.2}. Such a behaviour forbids the existence of velocity
 moments at arbitrary orders and is, thus, clearly undesirable. It turns out that the whole shape of $s_{0}(z)$ for model (i) can be pretty well 
understood from an adiabatic approximation, which is carried out in the 
Appendix \ref{AppendixI}, where solutions of arbitrary scaling 
exponents are 
constructed using adequate time reparametrizations and dilations of the 
deterministic solution of scaling exponent $z_{0}$. The validity of
this approximation for the model (i) is somehow  obvious from Fig. \ref{Fig.3},
where it can be checked that instantons keep indeed almost the same shape, even for quite sizable variations of $z$. Its failure 
in models (ii) and (iii) is probably due to too strong 
deformations of the solutions as $z$ increases, again suggested by Figs.
\ref{Fig.4} and \ref{Fig.5}. The full non-linear treatment of the problem proposed
in this paper was however necessary to reach this quite fortunate
 conclusion. \\

\section{The effect of quadratic fluctuations}
\label{fluct}
\subsection{Formal considerations}
\label{fluctformal}
In order to compute the first order (in $\Gamma$) correction to the action
per unit cascade step ($s_{1}(z)$ in the expression (\ref{pred_Pn})) of the 
density of probability $P_{n}(z)$, we have to expand the MSR action up to
quadratic order in fluctuations $\vectd{\delta}{b}$ around the extremal
trajectory $\vect{b}^{0}(\tau )$ of scaling exponent $z$ and then sum over
them in a way which will be explained below. Since typical fluctuating 
paths are not differentiable but rather behave as Wiener paths with derivatives 
of the order $1\over 2$, we shall stick to time-discretized expressions in all
the following manipulations of the path integral. For the sake of clarity, the
superscript 0 referring to the extremal trajectory in the previous Section
will be taken away, whereas $\vect{b}_{i}$,  $\vectg{\theta}_{i}$ will 
be short-hand notations for $\vect{b}^{0}(\tau_{i}=i\,\Delta \tau )$, 
$\vectg{\theta}^{0}(\tau_{i}=i\,\Delta \tau )$, where $\delta \tau$ is the
(small) time step used in the discretization.\\

We start from (\ref{pathint}) and the representation (\ref{MSRaction1})
of the effective action $S[\vect{b}, \vect{p}]$, expand it to quadratic order
in both fluctuations $\vectd{\delta}{p}$ and $\vectd{\delta}{b}$ and
then integrate out the fluctuations of the auxiliary field. To begin, the
time $\tau_{f}$ during which we let the system evolve will be equal to the
time $nT$ needed by the ideal instanton to perform $n$ steps along the
shell axis. The ideal initial and final configurations $\vect{b}_{in}=\vect{b}
_{0}$ and $\vect{b}_{f}=\vect{b}_{N}$ describe then a pulse centered 
successively around the shells of index $0$ and $n$. To quadratic order
in deviations from the extremal trajectory, the probability of joining the
perturbed endpoints $\vect{b}_{in}=\vect{b}_{0}+\vectd{\delta}{b}_{0}$
and $\vect{b}_{f}=\vect{b}_{N}+\vectd{\delta}{b}_{N}$ in the time
$\tau_{f}$ take the following expression~:
\begin{equation}
\label{prob_fluct}
P(\vect{b}_{0}+\vectd{\delta}{b}_{0},0;\vect{b}_{N}+\vectd{\delta}{b}_{N}, \tau_{f})={\e}^{-ns_{0}(z)/\Gamma}\int {\cal D} \vectd{\delta}{b}\,
\exp\left( -\delta S[\vectd{\delta}{b}]-
\delta ^{2}S[\vectd{\delta}{b}]\right)\,,
\end{equation}
where the measure of integration is defined as 
\begin{equation}
{\cal D} \vectd{\delta}{b}=\left(\frac{1}{2\pi \Gamma \Delta \tau }
\right)^{\frac{dN}{2}}\, \frac{1}{(\vect{b}_{0}\cdot\vect{b}_{0})^{d/2}}
\frac{1}{\mid \mbox{det}B_{0}\mid }
\prod _{i=1}^{N-1}
\left[\frac{d\vectd{\delta}{b}_{i}}{(\vect{b}_{i}\cdot\vect{b}_{i})
^{d/2}\mid \mbox{det}B_{i}\mid }\right]\,,
\end{equation}
the linear variation $\delta S$ reduces to the boundary term
\begin{equation}
\delta S[\vectd{\delta}{b}]=\frac{1}{\Gamma} 
\{ \vectg{\theta}_{N-1}\cdot\vectd{\delta}{b}_{N}
-\vectg{\theta}_{0}\cdot\vectd{\delta}{b}_{0}\}\,,
\end{equation}
and the quadratic one reads
\begin{equation}
\label{action_fluct1}
\delta ^{2}S[\vectd{\delta}{b}]=\frac{\Delta \tau }{2\Gamma}
\sum_{i=0}^{N-1} \left[(\vect{b}_{i}\cdot\vect{b}_{i})^{-1/2}\,B_{i}^{-1}\,
(\frac{\vectd{\delta}{b}_{i+1}-\vectd{\delta}{b}_{i}} {\Delta \tau}
-{\cal A}_{i}\vectd{\delta}{b}_{i})\right]^{2}
+\vectd{\delta}{b}_{i}.\,{\cal V}_{i}\,\vectd{\delta}{b}_{i}\,.
\end{equation}
The drift- and potential- terms in eq.(\ref{action_fluct1}) are found to be 
given by the following relations
\begin{equation}
{\cal A}_{i}\vectd{\delta}{b}_{i}={\cal M}_{i}\vectd{\delta}{b}_{i}
+(\vectd{\delta}{b}_{i}\cdot\partial_{\vect{b}_{i}})\left((\vect{b}_{i}\cdot\vect{b}_{i})
B_{i}\,^{t}\!B_{i}\vectg{\theta}_{i}\right)\,,
\end{equation}
and
\begin{equation}
\vectd{\delta}{b}_{i}.{\cal V}_{i}\vectd{\delta}{b}_{i}=
-\vectg{\theta}_{i}.(\vectd{\delta}{b}_{i}\cdot\partial_{\vect{b}_{i}})
{\cal M}_{i}\vectd{\delta}{b}_{i}-\frac{1}{2}\vectg{\theta}_{i}.
(\vectd{\delta}{b}_{i}\cdot\partial_{\vect{b}_{i}})^{2}
\left((\vect{b}_{i}\cdot\vect{b}_{i})B_{i}\,^{t}\!B_{i}\vectg{\theta}_{i}\right)\,,
\end{equation}
where $\cal M$ is, as in previous Sections, the Jacobian matrix of the
nonlinear kernel $\frac{\vect{N}[\vect{b}]}{(\vect{b}\cdot\vect{b})^{1/2}}$.\\

Our task is to perform explicitly the integration over fluctuations
$\vectd{\delta}{b}_{1}, \cdots, \vectd{\delta}{b}_{N-1}$ at intermediate
steps in the path integral (\ref{prob_fluct}). First it will be convenient to
get rid of the anisotropy of the ``mass'' tensor acting in the kinetic term of
$\delta^{2}S$ \cite{rem1} by switching to normalized fluctuating fields
 defined as
\begin{equation}
\vectd{\delta}{b}_{i}=(\vect{b}_{i}\cdot\vect{b}_{i})^{1/2}\,B_{i}\vect{h}_{i}\,.
\end{equation}
In this way the measure in the integral transforms into
$\frac{1}{(\vect{b}_{0}\cdot\vect{b}_{0})^{d/2}}
\frac{1}{\mid \mbox{det}B_{0}\mid }\,{\cal D}\vect{h}$
with
\begin{equation}
{\cal D} \vect{h}=\left(\frac{1}{2\pi \Gamma \Delta \tau }
\right)^{\frac{dN}{2}}\, \prod _{i=1}^{i=N-1}d\vect{h}_{i}\,.
\end{equation}
In performing this change of variables in $\delta^{2}S$, we must pay
attention to the fact that $\vectd{\delta}{b}_{i+1}-\vectd{\delta}{b}_{i}$,
as well as $\vect{h}_{i+1}-\vect{h}_{i}$, are potentially of order 
$\Delta \tau ^{1/2}$. One finds that, up to O($\Delta \tau ^{3/2})$ corrections
(negligible in the continuum limit), $\delta^{2}S$ becomes~:
\begin{equation}
\label{action_fluct2}
\delta ^{2}S[\vect{h}]=\frac{\Delta \tau }{2\Gamma}
\sum_{i=0}^{N-1} \left[Q_{i}^{1/2}\,(\frac{\vect{h}_{i+1}-\vect{h}_{i}} 
{\Delta \tau} +{\cal D}_{i}\frac{\vect{h}_{i+1}+\vect{h}_{i}}{2}
-{\cal A}'_{i}\vect{h}_{i})\right]^{2}
+\vect{h}_{i}.\,{\cal V}'_{i}\,\vect{h}_{i}\,,
\end{equation}
where
\begin{equation}
Q_{i}=\frac{1}{4}\,\left[1+\left(\frac{\vect{b}_{i+1}\cdot\vect{b}_{i+1}}
{\vect{b}_{i}\cdot\vect{b}_{i}}\right)^{1/2}\,^{t}\!B_{i+1}\,^{t}\!B_{i}^{-1}\right]\,
\left[1+\left(\frac{\vect{b}_{i+1}\cdot\vect{b}_{i+1}}
{\vect{b}_{i}\cdot\vect{b}_{i}}\right)^{1/2}\,B_{i}^{-1}\,B_{i+1}\right]\,,
\end{equation}
\begin{equation}
{\cal A}'_{i}=B_{i}^{-1}\,{\cal A}_{i}\,B_{i}\,, 
\end{equation}
\begin{equation}
{\cal V}'_{i}=(\vect{b}_{i}\cdot\vect{b}_{i})\,^{t}\!B_{i}\,{\cal V}_{i}\,B_{i}\,,
\end{equation} 
and
\begin{equation}
{\cal D}_{i}=\frac{2}{\Delta \tau}
\left[(\vect{b}_{i+1}\cdot\vect{b}_{i+1})^{1/2}B_{i+1}+
(\vect{b}_{i}\cdot\vect{b}_{i})^{1/2}B_{i}\right]^{-1}\,\left[(\vect{b}_{i+1}.
\vect{b}_{i+1})^{1/2}B_{i+1}- (\vect{b}_{i}\cdot\vect{b}_{i})^{1/2}B_{i}\right]\,.
\end{equation}
A few simplifications are now in order. On one hand we expand $Q_{i}$ as
$Q_{i}=1+\delta Q_{i} +O(\Delta \tau^{2})$, where
\begin{equation}
\delta Q_{i}=\left(\frac{\vect{b}_{i+1}\cdot\vect{b}_{i+1}}
{\vect{b}_{i}\cdot\vect{b}_{i}}\right)^{1/2}\,\frac{^{t}\!B_{i+1}\,^{t}\!B_{i}^{-1}
+B_{i}^{-1}\,B_{i+1}}{2}\,-1
\end{equation}
is of order $\Delta \tau$. Tracking $O(\Delta \tau )$ terms in $\delta^{2}S$,
we may replace $Q_{i}$ by 1 everywhere in (\ref{action_fluct2}), except for
the kinetic term $\frac{1}{2\Gamma}\frac{(\vect{h}_{i+1}-\vect{h}_{i}).
Q_{i}(\vect{h}_{i+1}-\vect{h}_{i})}{\Delta \tau}$, which according to 
standard computation rules \cite{LRT82} in path integrals can be reduced
to  $\frac{1}{2\Gamma}\frac{(\vect{h}_{i+1}-\vect{h}_{i})^{2}}{\Delta \tau} +\mbox{Tr}\,\delta Q_{i}$. On the other hand, exact ways of tracing out 
quadratic forms like $\delta^{2}S$ are available only once they are written
 in terms of the mid-point field $\frac{\vect{h}_{i+1}+\vect{h}_{i}}{2}$ rather than $\vect{h}_{i}$ (which amounts going back at this stage to the Stratonovich prescription). Again to order $\Delta \tau$, the substitution of
$\frac{\vect{h}_{i+1}+\vect{h}_{i}}{2}$ to $\vect{h}_{i}$ in 
(\ref{action_fluct2}) is harmless except in the cross-product term
$-\frac{1}{\Gamma}(\vect{h}_{i+1}-\vect{h}_{i})\cdot{\cal A}'_{i}\vect{h}_{i}$
which gets into 
$-\frac{1}{\Gamma}(\vect{h}_{i+1}-\vect{h}_{i})\cdot{\cal A}'_{i}
\displaystyle{\frac{\vect{h}_{i+1}+\vect{h}_{i}}{2}}
+ \frac{\Delta \tau}{2}\,\mbox{Tr}{\cal A}'_{i}$.\\

Putting things together, we arrive at the following expression for
the transition probability in the neighborhood of the instanton~:
\begin{equation}
\label{prob_fluct2}
P(\vectd{\delta}{b}_{in}\rightarrow \vectd{\delta}{b}_{f}, \tau_{f})=
\mid \mbox{det}\frac{\partial \vect{h}}{\partial 
\vect{b}}(\tau_{f})\mid \,{\e}^{-S_{0}(\tau_f)\Gamma}\,
{\e}^{-{\cal I}_{1}(\tau_{f})}\,{\e}^{\frac{1}{\Gamma}\{
\vectg{\theta}_{N-1}\cdot\vectd{\delta}{b}_{f}-\vectg{\theta}_{0}\cdot\vectd
{\delta}{b}_{in}\}}\, Z[\vect{h}_{in}\rightarrow \vect{h}_{f},\tau_{f}]\,,
\end{equation}
where we defined 
\begin{equation}
\label{prefact1}
{\cal I}_{1}(\tau_{f})=\frac{1}{2} \sum_{i=0}^{N-1}\mbox{Tr}\,\delta Q_{i}
+\frac{\Delta \tau}{2}\sum_{i=0}^{N-1} \mbox{Tr}\,{\cal A}_{i}
-d A\tau_{f}\,,
\end{equation}
and the reduced path integral $Z[\vect{h}_{in} \rightarrow 
\vect{h}_{f},\tau_{f}]$ as
\begin{equation}
\label{def_reducedpathint}
Z[\vect{h}_{in} \rightarrow \vect{h}_{f},\tau_{f}]=\int _{\vect{h}_{0}=
\vect{h}_{in}}^{\vect{h}_{N}=\vect{h}_{f}} {\cal D}\vect{h}\,
\exp -\delta^{2}S[\vect{h}]\,.
\end{equation}
with
\begin{equation}
\label{action_fluct3}
\delta ^{2}S[\vect{h}]=\frac{\Delta \tau }{2\Gamma}
\sum_{i=0}^{N-1} \left[\frac{\vect{h}_{i+1}-\vect{h}_{i}} 
{\Delta \tau} -{\cal B}_{i}\frac{\vect{h}_{i+1}+\vect{h}_{i}}{2}\right]^{2}
+\frac{\vect{h}_{i+1}+\vect{h}_{i}}{2}.\,{\cal V}'_{i}\,
\frac{\vect{h}_{i+1}+\vect{h}_{i}}{2}\,,
\end{equation}
and 
\begin{equation}
{\cal B}_{i}={\cal A}'_{i}-{\cal D}_{i}\,.
\end{equation}
One easily shows that in the $\Delta \tau \rightarrow 0$ limit
\[\sum_{i} \mbox{Tr}\delta Q_{i} \rightarrow \mbox{Tr}
\left[\mbox{ln}\frac{(\vect{b}_{N}\cdot\vect{b}_{N})^{1/2}B_{N}}
{(\vect{b}_{0}\cdot\vect{b}_{0})^{1/2}B_{0}}\right]=d\,A\tau_{f}\,,\]
since the periodicity of the instanton implies that $\mbox{det}
B_{N}=\mbox{det}B_{0}$, while 
\[\Delta \tau \sum_{i}\mbox{Tr}\,{\cal A}_{i}\rightarrow 
\int_{0}^{\tau_{f}} \{\partial_{\vect{b}}.[(\vect{b}\cdot\vect{b})B\,^{t}\!B
\vectg{\theta}]-\frac{\vect{N}[\vect{b}]\cdot\vect{b}}
{(\vect{b}\cdot\vect{b})^{3/2}}\}\,d\tau\,.\]
However the ensuing expression for ${\cal I}_{1}(\tau_{f})$ is not yet
complete to order $O(\Gamma^{0})$. This is because we discarded the $O(\Gamma)$ Ito drift-term in our computation of extremal trajectories.
The small deviations induced by this term may be neglected in ${\cal I}_{1}
(\tau_{f})$ and $-\mbox{ln}\, Z[\vect{h}_{in} \rightarrow 
\vect{h}_{f},\tau_{f}]$ which are already first order corrections in a 
$\Gamma$-expansion but they must be taken care of in the zeroth order term. Rather than $n\displaystyle{\frac{s_{0}(z)}{\Gamma}}$, 
it should read $S_{\Gamma}[\vect{b}^{0}_{\Gamma}]$, where the index $\Gamma$ denotes a quantity or field evaluated in the presence of the 
Ito-term specified in (\ref{Itokernel}). To first order in $\Gamma$, we can 
take advantage of
 the extremum property of $\vect{b}_{\Gamma}^{0}$ and write down 
\begin{eqnarray*}
S_{\Gamma}[\vect{b}_{\Gamma}^{0}]-S[\vect{b}^{0}]&=&
S_{\Gamma}[\vect{b}_{\Gamma}^{0}]-S_{\Gamma}[\vect{b}^{0}]+
S_{\Gamma}[\vect{b}^{0}]-S[\vect{b}^{0}]\\
&\approx & S_{\Gamma}[\vect{b}^{0}]-S[\vect{b}^{0}]=
\displaystyle{\frac{1}{\Gamma}\int _{0}^{\tau_{f}}\vectg{\theta}^{0}.
\frac{\vect{N}(\vect{b}^{0}]-\vect{N}_{\Gamma}[\vect{b}^{0}]}
{(\vect{b}^{0}\cdot\vect{b}^{0})^{1/2}}\,d\tau },
\end{eqnarray*}
 where the last identity just comes from the expression (\ref{MSRaction2}) of the action. Rearranging 
things under the assumption (satisfied by each of the particular models 
that we considered) that the entry $B_{jk}$ of the matrix $B$ involves
homogeneous monomials of degree $l$ built up only from components 
$C_{m}$ with $m\neq j$ (one has $l=0$ for model (i) and $l=2$ for
models (ii) and (iii) defined in Section \ref{def}), one finds
that ${\cal I}_{1}(\tau_{f})$ in (\ref{prob_fluct2}) should finally be
 understood as~:
\begin{equation}
\label{prefact2}
{\cal I}_{1}(\tau_{f})=\frac{(1-d)}{2}A\tau_{f}+\frac{(2l-3)}{4}
\int_{0}^{\tau_{f}} \vect{b}^{0}\cdot B\,^{t}\!B\vectg{\theta}^{0}\,d\tau\,.
\end{equation}

This preliminary work being done, the discussion will now concentrate on
the reduced path integral $Z[\vect{h}_{in} \rightarrow
\vect{h}_{f},\tau_{f}]$. We restrict first our attention to the case of fixed
endpoints $\vect{h}_{in}=\vect{h}_{f}=\vect{0}$. The instantons found in 
the previous Sections are physically satisfactory only if the quadratic
functional $\delta^{2}S[\vect{h}]$ is positive for all $\{\vect{h}_{i}\}$
such that $\vect{h}_{0}=\vect{h}_{N}=\vect{0}$ (we shall see later on that
 this is not a sufficient condition in the present problem). According to
standard results of functional analysis \cite{GF63}, positiveness of 
$\delta^{2}S[\vect{h}]$ is tantamount to the absence of points conjugate
 to the origin during the whole time interval $[0, \tau_{f}]$. Recall that the
definition of conjugate points goes as follows~: $d$ being the dimension of
the space (here the number of shells) we construct $d$ initial conditions
$(\vect{h}_{0}^{(\alpha)}, \vect{h}_{1}^{(\alpha)})$ such that
\[h_{0\beta}^{(\alpha)}=0,\quad \quad \quad h_{1\beta}^{(\alpha)}
=\Delta \tau \,\delta_{\alpha \beta}+O(\Delta \tau^{2})\,,\]
where $\beta$ is a shell index running as $\alpha$ between 1 and $d$ and
let them evolve under the Euler-Lagrange equations derived from
$\delta^{2}S[\vect{h}]$. The time $\tau_{i}=i\Delta \tau$ is said to be
 conjugate to the origin if the system formed by the $d$ vectors 
$\vect{h}_{i}^{(\alpha)}$ gets degenerate there. One can build a matrix
$U_{i}$ such that $U_{i}^{\alpha \beta}=h_{i\alpha}^{\beta}$, in terms of
which the initial conditions read
\begin{equation}
\label{initialcond}
U_{0}=0, \quad \quad \quad U_{1}=\Delta \tau +O(\Delta \tau^{2})\,,
\end{equation}
while $U_{i+1}$ (for $1\leq i\leq N-1$) is obtained from $U_{i-1}$ and
$U_{i}$ through a matrix Euler-Lagrange equation,  derived from
(\ref{action_fluct3}) and conveniently cast into the following form~:
\begin{equation}
\label{matrixEL}
(\frac{1}{\Delta \tau }+\frac{1}{2}\,^{t}\!{\cal B}_{i})P_{i}
-(\frac{1}{\Delta \tau }-\frac{1}{2}\,^{t}\!{\cal B}_{i-1})P_{i-1}
=\frac{1}{2}{\cal V}'_{i}\,\frac{U_{i}+U_{i+1}}{2}+
\frac{1}{2}{\cal V}'_{i-1}\,\frac{U_{i-1}+U_{i}}{2}\,,
\end{equation}
 where
\begin{equation}
\label{matrixmomentum}
P_{i}=\frac{U_{i+1}-U_{i}}{\Delta \tau}-{\cal B}_{i}\,\frac{U_{i}+U_{i+1}}
{2}\,,
\end{equation}
can be seen as a matrix momentum. In the absence of conjugate points, 
$\mbox{det}\,U_{i}$ never vanishes except at the origin and one gets the 
following simple expression for the reduced path integral, provided the
$\Delta \tau \rightarrow 0$ limit is ultimately taken~:
\begin{equation}
\label{Z_fixed}
Z[\vect{h}_{in}=\vect{0} \rightarrow \vect{h}_{f}=\vect{0},\tau_{f}]
=\left(\frac{1}{2\pi \Gamma}\right)^{d/2}\,
\frac{1}{\sqrt{\mbox{det}U(\tau_{f})}}\,.
\end{equation}
Details on this result, which may be found in many textbooks on path
integrals \cite{DR94,S81}, are provided in the Appendix \ref{AppendixII}. 
A very nice feature of the proof
of the connection of the positiveness of $\delta^{2}S[\vect{h}]$ with the
absence of conjugate points is that it provides also an efficient way for
computing the reduced path integral with $\vect{h}_{f}$ arbitrary
(but still small naturally). Indeed the main idea consists in adding to
$\delta^{2}S[\vect{h}]$ a boundary term of the form
\[
\delta^{2}S'=-\frac{1}{2\Gamma} \sum_{i=0}^{N-1}\left(\vect{h}_{i+1}\cdot
W_{i+1}\vect{h}_{i+1}-\vect{h}_{i}\cdot W_{i}\vect{h}_{i}\right)\,,
\]
where $W$ is a symmetric matrix at all times and then selecting the
right $W$ such that $\delta^{2}S+\delta^{2}S'$ becomes a perfect square.
In order to achieve this task, $W_{i}$ must be a solution of a matrix 
Ricatti equation (see again the Appendix B for details) which, upon the
substitution of a new unknown matrix $U_{i}$ defined implicitely by
the relation (for $0\leq i \leq N-1$)~:
\begin{equation}
\label{lienWU}
\frac{1}{2}(W_{i}U_{i}+W_{i+1}U_{i+1})=\frac{U_{i+1}-U_{i}}{\Delta \tau}
-{\cal B}_{i}\,\frac{U_{i}+U_{i+1}}{2}\,,
\end{equation}
is found to be nothing but the matrix Euler equation (\ref{matrixEL}). It
follows that the result (\ref{Z_fixed}) may be extended to the case of
an arbitrary final configuration $\vect{h}_{f}$ as 
\begin{equation}
\label{Z_mobile}
Z[\vect{h}_{in}=\vect{0} \rightarrow \vect{h}_{f},\tau_{f}]
=\left(\frac{1}{2\pi \Gamma}\right)^{d/2}\,
\frac{1}{\sqrt{\mbox{det}U(\tau_{f})}}\, \exp -\frac{1}{2\Gamma}
\vect{h}_{f}\cdot W(\tau_{f})\vect{h}_{f}\,,
\end{equation}
where $U(\tau_{f})$ and $W(\tau_{f})$ are to be computed by letting both
matrices evolve from $\tau =0$ to $\tau =\tau_{f}$ according to 
(\ref{matrixEL}) and (\ref{lienWU}). There are some subtleties about the 
choice of initial conditions and proper counting of the number of unknowns
whose discussion we prefer to relegate in the Appendix \ref{AppendixII}.

\subsection{A physical definition of $s_{1}(z)$}
\label{fluctdefs1}
We are now in a good position to compute the next to leading order term
$s_{1}(z)$ in the expansion of $-\lim_{n\rightarrow +\infty}\frac{1}{n}
 \log P_{n}(z)$ in powers of $\Gamma$. We shall obtain an estimate for
$P_n(z)$ by summing over all the trajectories which lead to the same
growth of the pulse as the ideal instanton $\vect{b}^{0}(\tau )$ after $n$
cascade steps. In the small $\Gamma$ limit, all
statistically relevant trajectories remain close to $\vect{b}^{0}(\tau )$
and we may define unambiguously  their ``arrival'' time at the shell of 
index $n$ as the first time $\tau_{n}$ such that
\begin{equation}
\vect{b}(\tau_n)=\vect{b}^{0}(\tau_{n}^{0}=nT)+\vectd{\delta}{b}'\,,
\end{equation}
where $\vectd{\delta}{b}'$ reduces to a linear combination of stable
``irrelevant'' modes $\vectg{\Phi}_{ir}(\tau_{n}^{0})$ for $i\geq 3$.
Up to multiplicative factors growing at most algebraically with $n$, we
can then write $P_{n}(z)$ as the following integral over the arrival time
and the position of the endpoint
\begin{equation}
\label{finalP_nz1}
P_{n}(z) \approx \int P(\vect{b}^{0}(0) \rightarrow 
\vect{b}^{0}(\tau_{n}^{0})+\vectd{\delta}{b}',\tau_{n}) 
\,\delta (\vectg{\Phi}_{1l}(\tau_{n}^{0}).
\vectd{\delta}{b}')\,\delta (\vectg{\Phi}_{2l}(\tau_{n}^{0}).
\vectd{\delta}{b}')\,d\tau_n\,d\vectd{\delta}{b}'\,.
\end{equation}
The density of probability in the right hand side of this expression is known from (\ref{prob_fluct2}) and (\ref{Z_mobile}), where $\tau_f$
should be taken equal to $\tau_n$ and $\vectd{\delta}{b}_{f}$ 
equal to $\vectd{\delta}{b}=\vect{b}(\tau_n)-\vect{b}^{0}(\tau_n)$. 
Calling  $\delta \tau=\tau_{n}-\tau^{0}_{n}$ the time delay, we get the following relation between $\vectd{\delta}{b}$ and $\vectd{\delta}{b}'$,
 valid up to $O(\delta \tau^{3})$ corrections~:
\begin{equation}
\label{lienfluct}
\vectd{\delta}{b}=\vectd{\delta}{b}'+\vect{b}^{0}(\tau_{n}^{0})-
\vect{b}^{0}(\tau_n)=\vectd{\delta}{b}'-\delta \tau \frac{d\vect{b}^{0}}
{d\tau}(\tau_n)+\frac{1}{2}\delta \tau^{2} \frac{d^{2}\vect{b}^{0}}
{d\tau^{2}}(\tau_{n}^{0})\,.
\end{equation}
Note that $\delta \tau$ scales typically like $\sqrt{\Gamma}$ in the 
semi-classical limit. We may thus, to leading order,  replace $\tau_n$
by $\tau_{n}^{0}$ and identify $\vectd{\delta}{b}$ with
$\vectd{\delta}{b}'-\delta \tau \,\frac{d\vect{b}^{0}}{d\tau}(\tau_{n}^{0})$ 
everywhere in the integrand of the right hand side of (\ref{finalP_nz1}),
except in $\frac{1}{\Gamma }(S_{0}(\tau_n)+\vectg{\theta}^{0}
(\tau_n).\delta
\vect{b})$ (the combination appearing in the exponential prefactor of
(\ref{prob_fluct2})) which deserves more care. It follows from the 
definition of the action that (again up to $O(\delta \tau^{3})$ corrections)
\begin{equation}
S_{0}(\tau_n)=S_{0}(\tau_{n}^{0})+\delta \tau \,\frac{(\vectg{\xi}^{0}
(\tau_{n}))^{2}}{2}-\frac{1}{2} \delta \tau^{2}\,\vectg{\xi}^{0}(\tau_n^{0}).
\frac{d\vectg{\xi}^{0}}{d \tau}(\tau_n^{0})\,,
\end{equation}
and from (\ref{lienfluct}) (together with the condition 
$\vectg{\theta}^{0}(\tau_{n}^{0})\cdot\vectd{\delta}{b}'=0$) that~:
\begin{equation}
\vectg{\theta}^{0}(\tau_n)\cdot\vectd{\delta}{b}=\delta \tau\, \frac{d\vectg{\theta}^{0}}{d\tau}(\tau_n^{0})\cdot\vectd{\delta}{b}'
-\delta \tau \,\vectg{\theta}^{0}(\tau_n)\cdot \frac{d\vect{b}^{0}}
{d\tau}(\tau_n)+\frac{1}{2}\delta \tau^{2} \vectg{\theta}^{0}(\tau_{n}^{0}).
\frac{d^{2}\vect{b}^{0}}{d\tau^{2}}(\tau_{n}^{0})\,,
\end{equation}
When summing these two equations, linear terms in $\delta \tau$ disappear as expected and we are left after some rearrangements with
the remainder, of quadratic order in fluctuations~:
\begin{equation}
\frac{1}{\Gamma}\{S_{0}(\tau_{n})+\vectg{\theta}^{0}(\tau_{n}).
\vectd{\delta}{b}\}=+\frac{1}{2}\delta \tau^{2}\frac{d\vectg{\theta}^{0}}
{d\tau}(\tau_{n}^{0})\cdot\frac{d\vect{b}^{0}}{d\tau}(\tau_{n}^{0})
+\delta \tau \,\frac{d\vectg{\theta}^{0}}{d\tau}(\tau_{n}^{0})\cdot
 \vectd{\delta}{b}\,,
\end{equation}
where we set $\vectd{\delta}{b}\equiv \vectd{\delta}{b}'-
\delta \tau\,\displaystyle{ \frac{d\vect{b}^{0}}{d\tau}}(\tau_{n}^{0})$, so 
that the time delay $\delta \tau$ is simply expressed in terms of
 $\vectd{\delta}{b}$ 
as $\delta \tau =-\vectg{\Phi}_{2l}(\tau_{n}^{0})\cdot\vectd{\delta}{b}$.\\

The end of our theoretical considerations is reached with the following
expression for $P_n(z)$~:
\begin{equation}
\label{finalP_nz2}
P_{n}(z) \approx {\e}^{-ns_{0}(z)/\Gamma}\,\frac{{\e}^{-{\cal I}_{1}(\tau_{n}^{0})}}
{\sqrt{\mbox{det}U(\tau_{n}^{0})}}\,\int
\frac{{\e}^{-\frac{1}{2\Gamma}\vect{h}\cdot\tilde{W}(\tau_{n}^{0})
\vect{h}}}{(2\pi \Gamma)^{d/2}}
 \delta (\vectg{\Phi}'_{1l}(\tau_{n}^{0})\cdot\vect{h})\,d\vect{h}\,,
\end{equation}
where $\vectg{\Phi}'_{1l}(\tau )=(\vect{b}^{0}\cdot\vect{b}^{0})^{1/2}\,^{t}\!B
\vectg{\Phi}_{1l}(\tau )$ (so that $\vectg{\Phi}'_{1l}\cdot\vect{h}
=\vectg{\Phi}_{1l}\cdot\vectd{\delta}{b}$), $\tilde{W}=W+\Delta W$,
with $W$ defined in the last subsection and 
\begin{equation}
\label {correctionW}
\vect{h}.\Delta W\,\vect{h}=-2(\vectg{\Phi}'_{2l}\cdot\vect{h})\,(\vect{b}^{0}.
\vect{b}^{0})^{1/2}\frac{d\vectg{\theta}^{0}}{d\tau}\cdot B\vect{h}
+(\vectg{\Phi}'_{2l}\cdot\vect{h})^{2}\,\frac{d\vectg{\theta}^{0}}{d\tau}\cdot
\frac{d\vect{b}^{0}}{d\tau}\,.
\end{equation}
We see that the condition of positiveness of the matrix $U$ must be
supplemented by the condition of positiveness of the restriction 
$\tilde{W}_{1}$ of $\tilde{W}$ to the $(d-1)$-dimensional space
orthogonal to $\vectg{\Phi}'_{1l}$, in order to make the instantons
found in the preceding Section physically meaningful. Provided these
two requirements are met, $s_{1}(z)$ is obtained as 
\begin{equation}
\label{expr_s1z}
s_{1}(z)=\lim_{n\rightarrow +\infty} \frac{1}{n}\{{\cal I}_{1}(nT)
+\frac{1}{2}\mbox{ln det}U(nT)+\frac{1}{2}\mbox{ln det} \tilde{W}_{1}
(nT)\}\,,
\end{equation}
which is the main result of this Section.\\

Tracing  back  all the steps leading to (\ref{expr_s1z}), one could object
to our starting point (\ref{finalP_nz1}) the fact that fluctuations of the
initial endpoint are not taken into account. This could be done at the 
expense of rather more cumbersome formulae for the reduced path 
integral $Z[\vect{h}(0)\rightarrow \vect{h}(\tau )]$ when both $\vect{h}(0)$ and $\vect{h}(\tau )$ do not vanish. However, we believe on physical grounds that
the blowing-up associated with the instanton washes out any influence of the
fluctuations at large scales on the part of the action scaling linearly with 
the number of steps $n$. Therefore the expression (\ref{expr_s1z}) should be
exact.

\subsection{Practical implementing and results}
\label{fluctimplementing}

\begin{figure}[l]
\centerline{\psfig{file=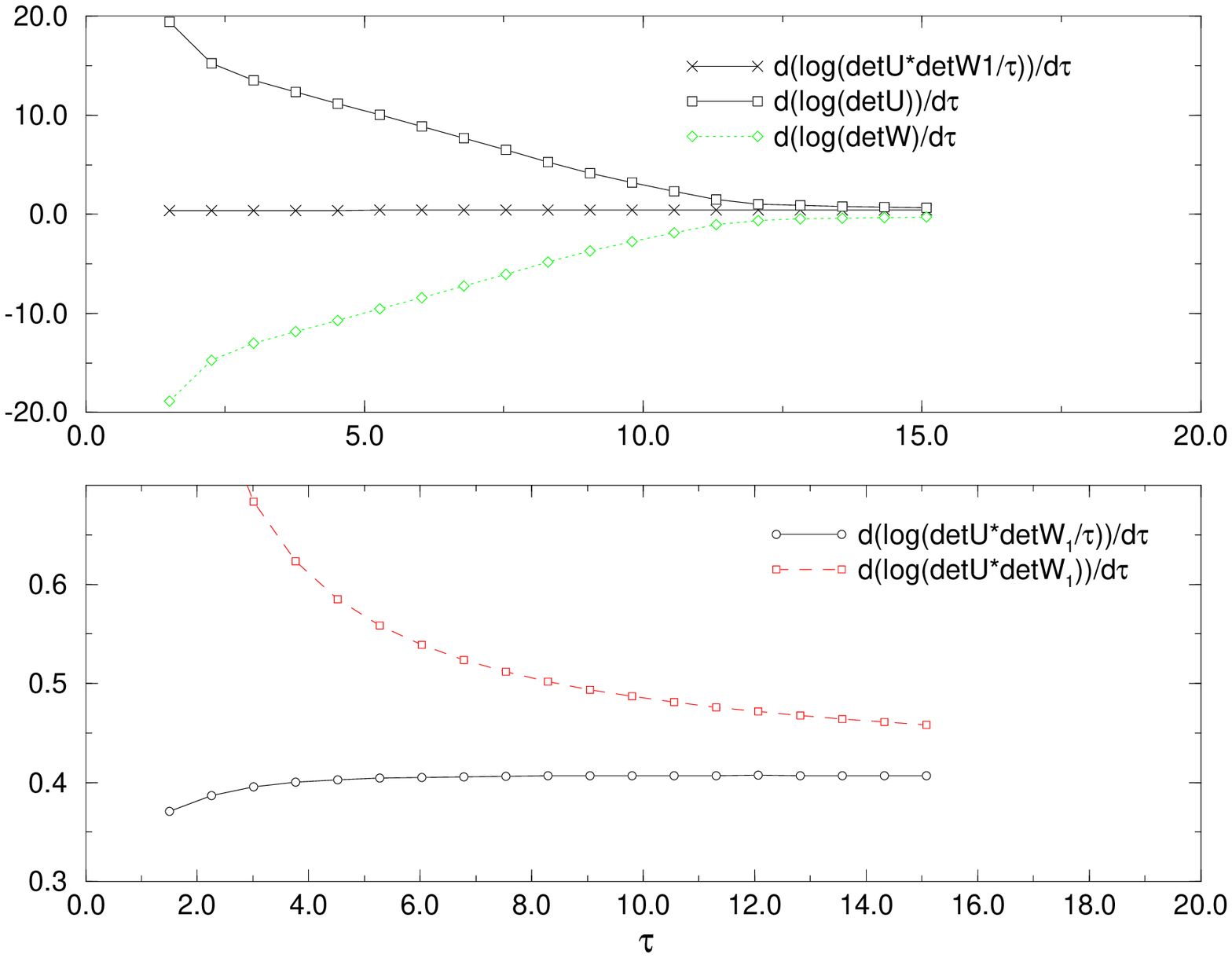}}
\caption{Test for the convergence of several relevant quantities entering the 
calculation of the function $S_1(z)$. We show the case of the model (ii) 
 at $z=0.8$.  }
\label{Fig.6}
\end{figure}

\begin{figure}[l]
\centerline{\psfig{file=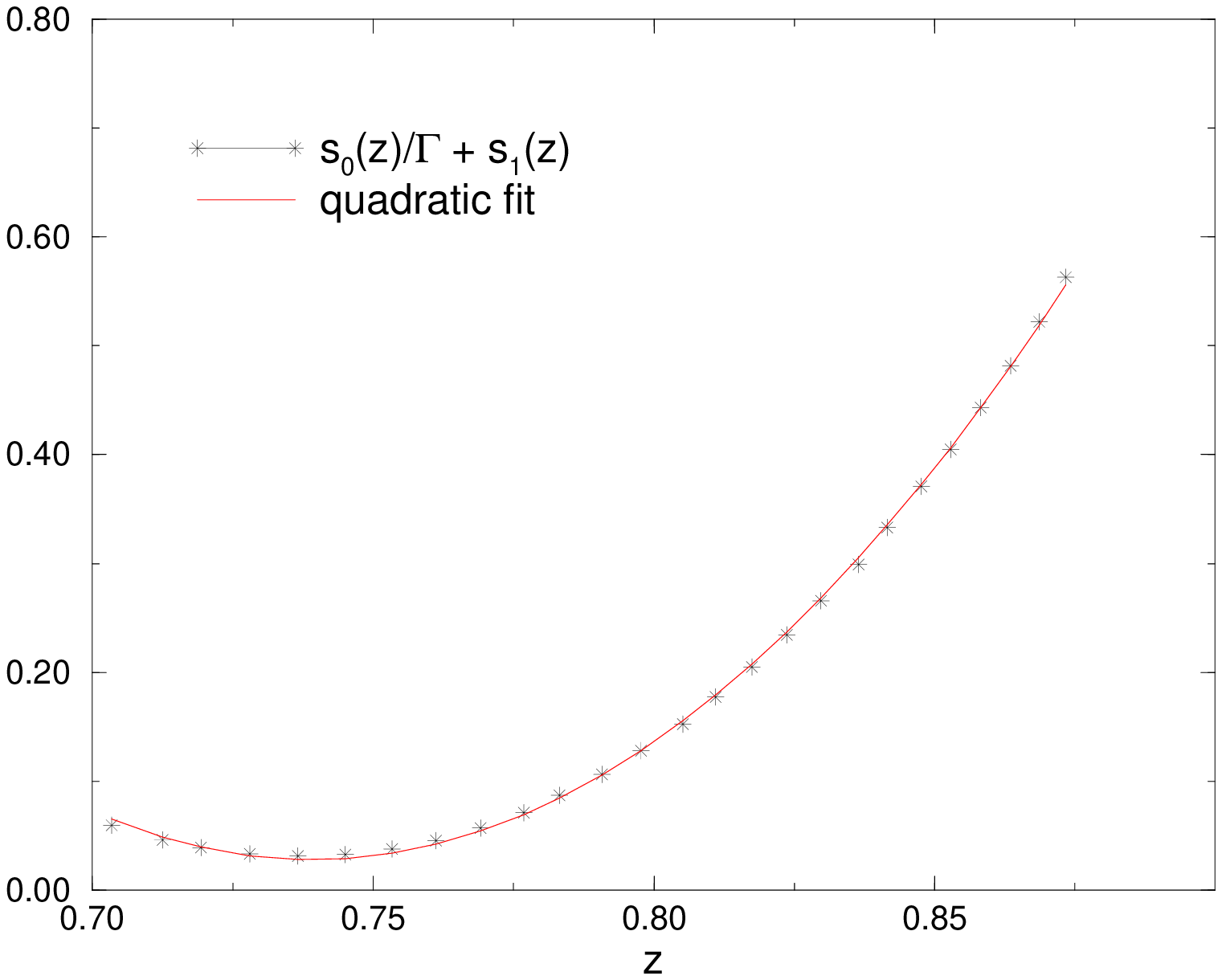}}
\caption{Graph (stars) of  $s_0(z)/\Gamma + s_1(z)$ for $\Gamma=0.58$ and model (ii). The parabolic fit $S_{quad}=a(z-z_{\star})^2$ yields $z_{\star}=0.74$
 and $a=30$.}
\label{Fig.7}
\end{figure}

The most difficult part of the computation of $s_{1}(z)$ lies in the
evaluation of $\det U(\tau )$ and $\det \tilde{W}_{1}(\tau )$ (as
defined in the preceding subsection), which requires a good control
of all the eigenvalues of these two matrices. Numerical instabilities could
be avoided for a time long enough to get a precise estimate of 
$\frac{d}{d\tau }\mbox{ln}\det U\tilde{W}_{1}$ by using the exact
expression of the matrix Euler equation (\ref{matrixEL}) and the
relation (\ref{lienWU}) between $W$ and $U$. In this problem, there
are two Goldstone modes associated with uniform rescaling and time
translation of the instanton~: as a consequence, the matrix 
$(\vect{b}^{0}\cdot\vect{b}^{0})^{1/2}BU(\tau )$ (resp.  
$(\vect{b}^{0}\cdot\vect{b}^{0})^{-1}\,{^{t}\!B^{-1}}\tilde{W}B^{-1}$) is
expected to have two eigenvalues scaling like $\tau $ (resp. like
$1/\tau $) (in order to obtain the simplest transcription of these
symmetries, one has to go back to the original fluctuation field
$\vectd{\delta}{b}=(\vect{b}^{0}\cdot\vect{b}^{0})^{1/2}B\vect{h}$ and 
the change of variable influences eigenvalues of $U$ and $\tilde{W}$). 
When restricting $\tilde{W}$ to the $(d-1)$-dimensional space orthogonal
to $\vectg{\Phi}'_{1l}$ ($\equiv (\vect{b}^{0}\cdot\vect{b}^{0})^{1/2}
B\vectg{\Phi}_{1l}$), one looses one of the eigenvalues scaling like
$1/\tau $, so that there remains an algebraic factor $\sqrt{\tau }$ in the
 product $\sqrt{\det U(\tau )}\sqrt{\tilde{W}_{1}(\tau )}$, which we had to
take away by hand in order to make more conspicuous the leading 
exponential growth of this quantity. To give an idea of the accuracy of our
 procedure and confirm the soundness of the intricate formula
(\ref{expr_s1z}) that was proposed for $s_{1}(z)$, we show in Fig. \ref{Fig.6}
 the behaviour of
various relevant quantities in the case of model (ii), and for a moderate
scaling exponent $z=0.8$. It is observed in the picture on the top that the
logarithmic derivatives of $\det U$ and $\det W$ exhibit a linear 
behaviour with almost opposite slopes. It can be shown by considering 
simpler and exactly solvable models for quadratic fluctuations without
inter-shell couplings, that this strange effect mostly reflects the stiff
variations suffered by the noise variance on any shell, as the latter moves
back from the leading edge of the instanton to its rear end. The argument is
presented in the Appendix \ref{AppendixIII}, since it may help the reader to
 get a feeling
for the order of magnitudes at play in both $U$ and $W$ matrices. \\
It could be tempting at this level to approximate $s_{1}(z)$ as
\begin{equation}
\label{simple_s1z}
\lim_{\tau \rightarrow \infty}\frac{d}{d\tau }\{{\cal I}_{1}+\frac{1}{2}
\det \mbox{ln}WU\},
\end{equation}
This expression would come out if, without great physical justification, it were decided to sum over all positions of the final endpoint in the path
 integral formulation in order to estimate the volume of the basin of 
attraction of the instanton. We found however that the matrix momentum
 $WU$ develops invariably a negative eigenvalue after some time, so that
the projection onto the restricted phase space introduced in the previous
subsection is a necessary step for restoring the statistical stability of the
instanton. Further, even before this instability occurs, there was found to
be a residual $\tau^{2}$ term in $\mbox{ln} \det WU$ which forbids any
reliable estimate for $s_{1}(z)$ to be deduced from (\ref{simple_s1z}). 
The picture at the bottom of Fig.  \ref{Fig.6} shows by contrast that the more
precise quantity $\det U \det \tilde{W}_{1}$ quickly settles to a
perfect exponential growth, once algebraic transients have been factored
 out. Note that the positiveness of both $U$ and $\tilde{W}_{1}$ could be
checked in any instance. \\
In all the models that we investigated, we found that the first order
correction $s_{1}(z)$ to the action takes an approximate parabolic shape
of positive or negative concavity, centered around a value of $z$ different
from $z_{0}$. In the case of the model (ii), the concavity of
$s_{1}(z)$ is  opposite to the one of $s_{0}(z)$ and the maximum of
 $s_{1}(z)$ is reached at a scaling exponent $z_{1} \approx 0.6$,
 significantly lower than $z_{0}=0.72$. 
This means that the minimum of the total action density $s(z) =
 s_{0}(z)/\Gamma+s_{1}(z)$ (for values of $\Gamma $ such that $s(z)$
remains concave as it should) is displaced toward the side of larger
exponents, just as a result of fluctuations. The trend is just the opposite
for the model (iii), where $s_{1}(z)$ presents this time the same concavity
 as $s_{0}(z)$ and a minimum on the left side of $z_{0}$. Fig. \ref{Fig.7}
 shows the graph of $s_{0}(z)/\Gamma+s_{1}(z)$ that is obtained for the model
(ii) and for the value $\Gamma = 0.58$, which we believe to be of some
 relevance for the GOY model (see Section \ref{discussion}).

\section{Comparison with numerical data on the sTatistics of coherent events
 in the GOY model}
\label{discussion}
\begin{figure}[p]
\centerline{\psfig{file=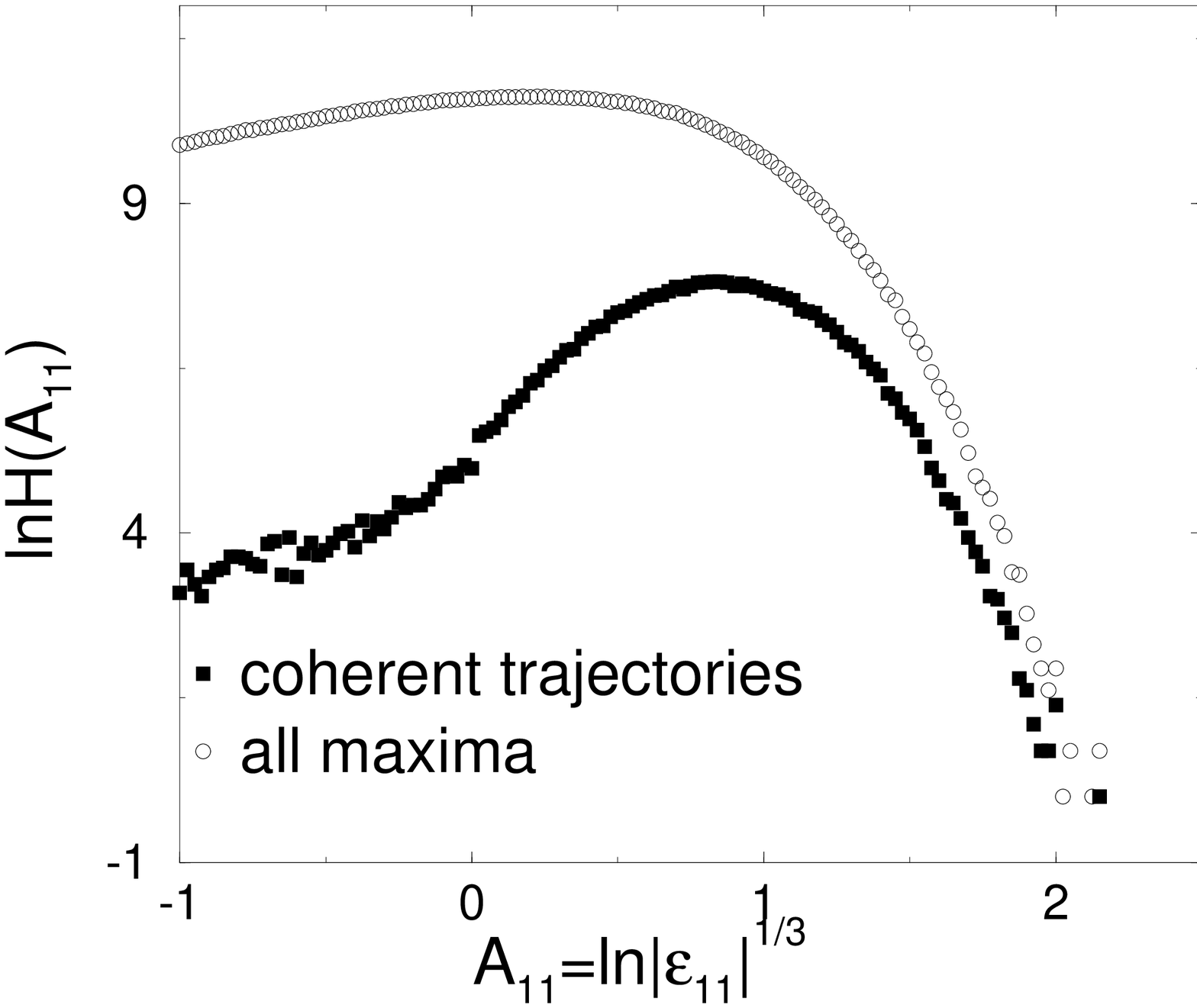,height=7cm,angle=-0}}
\caption{Histograms of the energy flux in log-log plot, involving all
relative maxima of $\epsilon_n$ or only those associated with coherent events.
The shell index $n=11$, the Reynolds number $Re=10^{8}$ and the number of
totls is $6\times 10^{4}$.
}
\label{Fig.8}
\end{figure}

\begin{figure}[p]
\centerline{\psfig{file=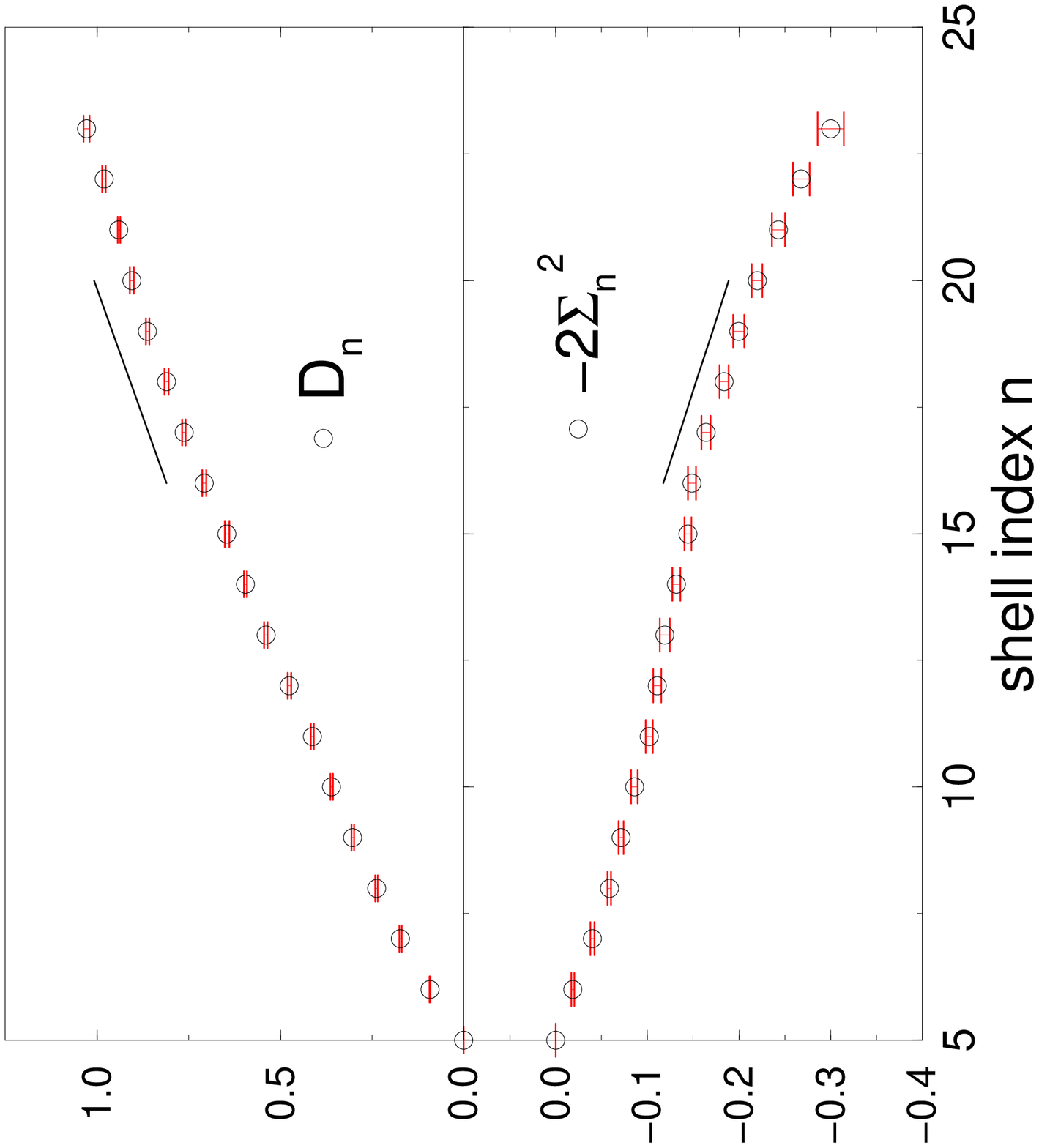,height=7cm,angle=-0}}
\caption{The two quantities $D_n$ and $-2\Sigma_n^2$ (encoding the Gaussian 
central part of the histogram of the growth variable $A_{n}-A_{n_{0}}$) 
{\em vs} $n$. The Reynolds number $Re=10^{9}$ and the dissipative shell has
 index $n_{d}=23$. The two pieces of straight line show the linear fits that
 were used to extract the values of $z_{\star}$ and $a$ in the pre-viscous
 range. 
} 
\label{Fig.9}
\end{figure}

The systems we have analyzed here were  
introduced to describe inertial singular structures of the GOY model. To
check their physical relevance, we first have to define a class of events
observed in full simulations of the GOY model which are likely to be the
best candidates for a description in terms of instantons.  
It is clear that relative maxima of the instantaneous energy flux
 $\epsilon_{n}(t)$ 
(with $\epsilon_n=k_{n}^{-2}{\rm Re}\{(1-\epsilon )Q^{2}b_{n-2}b_{n-1}b_n+
b_{n-1}b_{n+1}b_n\}$) are 
useful observables for tracking the passage of coherent structures across the
whole inertial range. But their total number is found to grow with $n$ as
$k_{n}^{2/3}$, due to the acceleration of time scales typical of the
Kolmogorov energy cascade. One may consider that they develop on tree-like
patterns in the $(n,t)$-plane, which are renewed at each turn-over of the large
scales (totls). We say that such trees provide a realization of the 
propagation of a
 coherent event from shell $n_{0}$ to shell $n>n_{0}$,
 whenever the $n-n_{0}$ nodes of the
 tree closest in time to their supposedly common ancestor on the shell of
lower index $n_{0}$ appear in the order of increasing shell index. In order
to discard too weak, and therefore irrelevant, events, we imposed that
$\epsilon_{n_{0}}$ be greater than half the mean energy flux.     
Figure \ref{Fig.8} shows the logarithm of the histogram of 
the logarithmic amplitudes $A_n=\ln|\epsilon_n|^{1/3}$ for all relative 
maxima on one hand and for the restricted class of coherent events defined above on the other hand, with $n_{0}=5$ (far enough from the forcing range) and 
$n=11$ (well in the inertial range). The Reynolds number 
of the simulation is $Re=10^8$. Statistics have been run over $6\times 10^4$ 
totls, and in average there are three 
``coherent lines'' for two 
totls. We note that the statistics of coherent events is very close to
 log-normal for $|\epsilon _n|\geq {\cal O}(1)$.\\
 The effective exponent $z$ of a coherent burst after $n-n_{0}$ cascade
steps is obtained via the relation $A_n=A_{n_0}+(n-n_0)(z-2/3)\ln 2$.
If anomalous scaling is preserved in the $Re\rightarrow \infty$ limit, the
pdf of scaling exponents $z$ should behave at large cascade lengths as
$P_{n}(z) \sim e^{-ns(z)}$,
where generically $s(z)$ will present a quadratic minimum at
some $z_{\star}$, with an expansion around $z_{\star}$ that we write as
 $s(z)=a(z-z_{\star})^{2}+\cdots $. The histogram of the variable
$A_{n}-A_{n_{0}}$ should consequently evolve, as $n-n_{0}$ increases, 
 towards a Gaussian shape, whose center $D_{n}$ and  
variance $\Sigma_{n}^{2}$ grow linearly with $n$ and
relate to $z_{\star}$ and $a$ as
\begin{equation}
D_{n}\sim n\ln 2 (z_{\star}-2/3), \quad \quad \quad \Sigma_{n}^{2}\sim
\frac{n}{2a}(\ln 2)^{2}.
\label{drift_var}
\end{equation}
The actual behaviours of $D_n$ and $-2\Sigma^2_n$, obtained from a very high
Reynolds number simulation ($Re=10^{9}$, which sets the dissipative scale
at the shell index $n_d=\frac{3}{4}\ln_{2}Re =23$) are plotted in 
Fig.~\ref{Fig.9}. Error bars were estimated by
varying the range of the quadratic fit to the logarithm of the histogram of
 $A_{n}-A_{n_{0}}$, as well as the domain of initial amplitudes $A_{n_{0}}$ 
used to construct this histogram. It appears that the two
graphs are rather far from simple straight lines : this is especially true
for the variance, whose graph from concave gets convex beyond the shell index
 15. From the investigation of lower Reynolds numbers $Re=10^{8}$ and
$Re=10^{7}$ we could deduce that this transition occurs at a shell index
$n_c$ always of the order of $n_{d}-8$ and defines a clear-cut boundary
between the inertial range and a surprisingly wide pre-viscous range.
We believe that the direct action of viscosity on intense bursts starts
to show up only at the shell index $n_{d}-3$, beyond which the local slope
 of the $D_{n}$-graph ceases to vary and points to a value of the average
growth exponent of coherent structures precisely equal to $z_0$. The fact
that $n_{c}$ lies rather far from $n_{d}$ means that the
cut-off imposed by viscosity exerts a long-range influence on the
statistics of the random environment
seen by a coherent structure. A decent
linear regime for both the drift and the variance is observed in the range
$15<n<21$, from which we get the
two estimates $a=29\pm 4$ and $z_{\star}=0.74 \pm 3.10^{-3}$.
Assuming that the fitting range $10<n<16$ provides the best clue to
 the asymptotic scaling of the inertial range, one gets the second
 set of values $a=45\pm 6$ and $z_{\star}=0.75\pm 3.10^{-3}$.
It appears that the physics of the previscous range is quite well reproduced
within our modelling (ii) of the incoherent background. By choosing
 $\Gamma =0.58$ (a value hopefully small enough to fall within the range
of validity of semiclassical approximations),  one obtains,
 as Fig. \ref{Fig.7} shows,
an almost perfect parabolic shape of $s(z)=-(S_{0}(z)/\Gamma
 +S_{1}(z))$, with a maximum reached at $z_{\star}=0.74$ and a curvature
$a=-2s''(z_{\star})=29$. However the parameter $\Gamma$ cannot be adjusted
 so as to account for the higher values of $a$ and $z_{\star}$ characterizing
 the inertial range. It would seem that in this range of scales it gets
 necessary to assume some bias in the incoherent fluctuations boosting
 the increase of the renormalized value of $z_{\star}$, while keeping the
 noise width small.\\

\section{Conclusion}
\label{conclusion}

We have developed a general scheme for computing numerically
self-similar instantons in scale invariant stochastic dynamical systems.
As concerns the physics of the GOY model, we believe that the bunch of results
presented here give a strong support to the relevance of an approach focusing
 from the outset on structures in order to understand intermittency and
treating the rest of the flow as a noise of weak amplitude. In particular
 the trend toward log-normal statistics of coherent structures is nicely
 recovered. The detailed study of various stochastic extensions of the
GOY model shows that the resulting pdf of scaling exponents of singular
strucures is very sensitive to the hypothesis made on the coupling of noise
to the velocity gradient fiels.\\
 We hope that our approach will be useful for attacking the 3D-Navier-Stokes
dynamics along similar lines, once an adequate decomposition of the flow into
coherent and incoherent parts will have been introduced. An application of
the method to the Kraignan's model of passive scalar advection formulated
on a lattice of shells has already been attempted \cite{BDDL99}. It has
given encouraging results with regard to the validity of a semi-classical 
analysis, even in situations where a small parameter (like $\Gamma $ in the
present problem) is missing.

\section{Acknowledgments}
 We thank L. Biferale, G. Falkovich, V. Hakim, V. Lebe\-dev, P. Muratore, 
D. Vandembroucq and R. Zeitak for useful discussions or  
suggestions at various stages of this work. J.-L. G. has been partly supported 
by the CNRS/CRTBT and by grants from the Israeli Science Foundation and the 
Minerva Foundation (Germany).\\

\newpage
\appendix
\section{}
\label{AppendixI}

We carry out in this Appendix the adiabatic approximation alluded to in
Section \ref{ExtSolimplementing}.  We look for self-similar instantons within the restricted
manifold of configurations of the type 
\begin{equation}
\label{trialinst}
\vect{b}(\tau )={\e}^{x(\tilde{\tau })}\vect{b}^{0}(\tilde{\tau }),
\end{equation}
where $\vect{b}^{0}(\tau )$ is the deterministic solution of scaling exponent
$z_{0}$, $\tilde{\tau }$ can be thought of as a ``proper'' time referring to the actual position of the pulse. The two variables $\tau (\tilde{\tau })$ and $x(\tilde{\tau})$ parameterize then local changes of speed and amplitudes 
of the pulse, which keeps the same shape as in the absence of noise. Note 
that if (\ref{trialinst}) is to represent a self-similar instanton of scaling
exponent $z\neq z_0$, $x(\tilde{\tau })$ must obey the constraint
\begin{equation}
\label{constraintx}
x(\tilde{\tau}+T_{0})-x(\tilde{\tau})=(z-z_{0})\log Q\,.
\end{equation}
We plug now the Ansatz (\ref{trialinst}) in the equation of motion
 (\ref{Ito_tau}) and project it onto the two directions $\vectg{\Phi}_{1r}=
\vect{b}^{0}$ and $\vectg{\Phi}_{2r}=\frac{d\vect{b}^{0}}{d\tau }$. We get
by doing so
\begin{eqnarray}
\label{reducedstoch1}
\frac{d\tilde{\tau}}{d\tau} \frac{dx}{d\tilde{\tau }}&=&\vectg{\Phi}_{1l}.
B\vectg{\xi}\,, \\
\frac{d\tilde{\tau}}{d\tau}-1&=&\vectg{\Phi}_{2l}\cdot B\vectg{\xi}\,.
\label{reducedstoch2}
\end{eqnarray}
If the other dimensions of the configuration space are neglected, (\ref{reducedstoch1}) and (\ref{reducedstoch2}) form a closed two-dimensional system, which may be rewritten as
\begin{eqnarray}
\frac{dx}{d\tilde{\tau}}&=&\zeta_{1}\,,\\
 1-\frac{d\tau}{d\tilde{\tau}}&=&\zeta_{2}\,,
\end{eqnarray}
where correlation functions of $\zeta_{1}$ and $\zeta_{2}$ read
\begin{equation}
\langle \zeta_{i}(\tilde{\tau})\zeta_{j}(\tilde{\tau}')\rangle
=\frac{d\tau}{d\tilde{\tau}} V_{ij}\,\delta (\tilde{\tau}-\tilde{\tau}')
\quad \quad 1\leq i,j, \leq 2,
\end{equation}
with 
\begin{equation}
V_{ij}=\vectg{\Phi}_{il}\cdot B\,^{t}\!B\vectg{\Phi}_{jl}\,.
\end{equation}
The Gaussian action density associated to one cascade step within this
 restricted stochastic system is given by
\begin{equation}
\tilde{s}=\frac{1}{2} \int_{0}^{T_{0}}d\tilde{\tau}\,
(\frac{d\tau }{d\tilde{\tau}})^{-1}\zeta_{i}(V^{-1})_{ij}\zeta_{j},
\end{equation}
Once expressed in terms of the diffusing variables $x(\tilde{\tau})$ and
 $\tau (\tilde{\tau})$,it becomes
\begin{eqnarray}
\tilde{s}&=&\frac{1}{2}\int_{0}^{T_{0}}d\tilde{\tau }\,
\{(\frac{d\tau}{d\tilde{\tau}})^{-1}\left[\dot{x}^{2}(V^{-1})_{11}
+2\dot{x}(V^{-1})_{12}+(V^{-1})_{22}\right]+\frac{d\tau}{d\tilde{\tau}}
(V^{-1})_{22}\nonumber \\
& &-\left[\dot{x}\,(V^{-1})_{12}+(V^{-1})_{22}\right]\}.
\end{eqnarray}
The extremization of $\tilde{s}$ with respect to
 $\frac{d\tau}{d\tilde{\tau}}$ leads to
\begin{equation}
\frac{d\tau}{d\tilde{\tau}}=\left[(\dot{x}^{2}(V^{-1})_{11}+2\dot{x}\,
(V^{-1})_{12}+(V^{-1})_{22})V_{22}\right]^{1/2},
\end{equation}
and to an effective action for the remaining variable $x$
\begin{eqnarray}
\tilde{s}_{eff}[x(\tilde{\tau})]&=&\int_{0}^{T_{0}}d\tilde{\tau }\,
\{\left[(\dot{x}^{2}(V^{-1})_{11}+2\dot{x}\,(V^{-1})_{12}+
(V^{-1})_{22})(V^{-1})_{22}\right]^{1/2}\nonumber \\
& &-\left[\dot{x}\,(V^{-1})_{12}+(V^{-1})_{22}\right]\}.
\end{eqnarray}
Assuming the coefficients $V_{11}$, $V_{12}$ and $V_{22}$ to be almost
constant inside the time interval $T_{0}$, one deduces an analytic expression for $s_{0}(z)$ from $\tilde{s}_{eff}$ by just replacing in the
integral $\dot{x}$ by $(z-z_{0})\log Q/T_{0}$ (which follows from 
(\ref{constraintx})). One gets in particular for large enough 
$z-z_{0}$, 
\begin{equation}
s_{0}(z)\sim (z-z_{0})\log Q\,\left\{((V^{-1})_{11}(V^{-1})_{22})^{1/2}-
(V^{-1})_{12}\right\},
\end{equation}
i.e, a linear behaviour as observed for the true solution of model (i).

\section{}
\label{AppendixII}

We derive the formal expression of the reduced path integral
$Z[\vect{0} \rightarrow \vect{h}_{f},\tau_{f}]$ given in
eq. (\ref{Z_mobile}) of the Section \ref{fluct} of the text. The proof is
presented in many textbooks on path integrals but, as far as we know, 
always using a continuous definition of time. This leaves some ambiguity
in the right equations that matrices $U$ and $W$ should obey, once time is
discretized for computing purposes. We found that this issue is crucial 
mostly for evaluating $W$ and preserving its symmetry properties. This is
why we feel it useful to show how every step of the proof given in the
continuum limit receives an exact transcription in the discrete time case.\\
We start from the quadratic functional~:
\begin{equation}
\label{quadra1}
\delta ^{2}S[\vect{h}]=\frac{\Delta \tau }{2\Gamma}
\sum_{i=0}^{i=N-1} \left\{\left[\frac{\vect{h}_{i+1}-\vect{h}_{i}} 
{\Delta \tau} -{\cal B}_{i}\frac{\vect{h}_{i+1}+\vect{h}_{i}}{2}\right]^{2}
+\frac{\vect{h}_{i+1}+\vect{h}_{i}}{2}.\,{\cal V}'_{i}\,
\frac{\vect{h}_{i+1}+\vect{h}_{i}}{2}\right\}\,,
\end{equation}
Upon the addition of the boundary term $-\frac{1}{2\Gamma}
\sum_{i=0}^{N-1}\{\vect{h}_{i+1}.
W_{i+1}\vect{h}_{i+1}-\vect{h}_{i}.W_{i}\vect{h}_{i}\}$, it becomes without 
any approximation~:
\begin{eqnarray}
\label{quadra2}
 \delta ^{2}\tilde{S}[\vect{h}]&=&\frac{\Delta \tau }{2\Gamma}
\sum_{i=0}^{i=N-1}\{ \left[\tilde{Q}_{i}^{1/2}\left(\frac{\vect{h}_{i+1}-\vect{h}_{i}} 
{\Delta \tau}-{\tilde{Q}_{i}}^{-1}({\cal B}_{i}-\frac{W_{i}+W_{i+1}}{2})\,
\frac{\vect{h}_{i+1}+\vect{h}_{i}}{2}\right)\right]^{2} \nonumber \\
& & \qquad +\frac{\vect{h}_{i+1}+\vect{h}_{i}}{2}.\,\tilde{\cal V}_{i}\,
\frac{\vect{h}_{i+1}+\vect{h}_{i}}{2}\}\,,
\end{eqnarray}
where
\begin{equation}
\label{defQtilde}
\tilde{Q}_{i}=1-\frac{\Delta \tau}{4}(W_{i+1}-W_{i})\,,
\end{equation}
and
\begin{equation}
{\tilde {\cal V}}_{i}={\cal V}'_{i}+^{t}\!{\cal B}_{i}{\cal B}_{i}-
\frac{W_{i+1}-W_{i}}{2}
-(^{t}\!{\cal B}_{i}-\frac{W_{i}+W_{i+1}}{2})\tilde{Q}_{i}^{-1}
({\cal B}_{i}-\frac{W_{i}+W_{i+1}}{2})\,.
\end{equation}
We see that $\delta^{2}\tilde{S}[\vect{h}]$ reduces to the time integral of
a single (positive) square, if and only if $W_{i}$ is such that for all 
$0\leq i \leq N-1$
\begin{equation}
\label{Ricatti}
{\cal V}'_{i}+^{t}\!{\cal B}_{i}{\cal B}_{i}-
\frac{W_{i+1}-W_{i}}{2}
-(^{t}\!{\cal B}_{i}-\frac{W_{i}+W_{i+1}}{2})\tilde{Q}_{i}^{-1}
({\cal B}_{i}-\frac{W_{i}+W_{i+1}}{2})=0\,.
\end{equation}
We note that (\ref{Ricatti}) forces $W_{i}$ to remain symmetric for all time,
provided that $W_{0}$ (arbitrary at this stage) is chosen to be such. In order to solve the above Ricatti matrix equation, one makes the following
change of matrix variable 
\begin{equation}
\label{defU}
({\cal B}_{i}+\frac{W_{i}+W_{i+1}}{2})(\frac{U_{i}+U_{i+1}}{2})
=\tilde{Q}_{i}\,\frac{U_{i+1}-U_{i}}{\Delta \tau}\,.
\end{equation}
>From the expression (\ref{defQtilde}) of $\tilde{Q}_{i}$, it is easily 
shown that
(\ref{defU}) is equivalent to the equation (\ref{lienWU}) quoted in the text.
Furthermore, by multiplying both sides of (\ref{Ricatti}) by 
$\frac{U_{i}+U_{i+1}}{2}$ on the right, one gets for $0\leq i\leq N-1$
\begin{equation}
\label{secondlienWU}
\frac{W_{i+1}U_{i+1}-W_{i}U_{i}}{\Delta \tau}=-^{t}\!{\cal B}_{i}\,
\frac{U_{i+1}-U_{i}}{\Delta \tau}+(^{t}\!{\cal B}_{i}{\cal B}_{i}+{\cal V}'_{i})
\frac{U_{i}+U_{i+1}}{2}\,.
\end{equation}
By half-summing the two relations yielded by (\ref{secondlienWU}) at
subsequent values $i-1$ and $i$ of the temporal index (with then $1\leq 
i\leq N-1$), and using (\ref{lienWU}) after noticing that $\frac{W_{i+1}U_{i+1}-W_{i-1}U_{i-1}}{2\Delta \tau}=\frac{(W_{i+1}U_{i+1}
+W_{i}U_{i})-(W_{i}U_{i}+W_{i-1}U_{i-1})}{2\Delta \tau}$, one can
eliminate $W$ and check that $U$ obeys the matrix Euler equation (\ref{matrixEL}), as promised in the text. \\

So far, we have proven that, as long as the matrix $U$ may be inverted,
 the positiveness of $\delta^{2}S$ is guaranteed, since
in that case the matrix $W$ exists at all times (from (\ref{defU})) and allows one to transform the initial quadratic form into the time integral of a 
single square. We show now how these matrices lead to a compact 
expression of $Z[\vect{h}_{0}=\vect{0}\rightarrow \vect{h}_{f},\tau_{f}]$.
We first note that (\ref{defU}) and (\ref{secondlienWU}) provide $2N$
relations for $2(N+1)$ unknowns $\{U_{0},\cdots ,U_{N}\}$, $\{W_{0},
\cdots ,W_{N}\}$. This gives much freedom in the choice of $W_{0}$
and $U_{0}$. In the particular case of $\vect{h}_{0}=\vect{0}$, it is
convenient to set $U_{0}=0$ and $W_{0}U_{0}=1$ (which should be 
understood as the limit as $\epsilon \rightarrow 0^{+}$ of $U_{0}=\epsilon$
and $W_{0}=\epsilon^{-1}$, so that $W_{0}$ is indeed symmetric). A quick
inspection of (\ref{defU}) and (\ref{secondlienWU}) reveals that $U_{i}$ 
and $W_{i}$ behave then respectively as $i\Delta \tau$ and $(i\Delta \tau)
^{-1}$ to leading order in $\Delta \tau$ for $1 \leq i \ll N$~. One has for
instance the exact result $W_{1}=(\Delta \tau )^{-1}-\frac{{\cal B}_{0}+
^{t}\!{\cal B}_{0}}{2}+(^{t}\!{\cal B}_{0}{\cal B}_{0}+{\cal V}'_{0})$. Recall
that the quantity we wish to estimate reads now~:
\begin{equation}
\label{Z_mobile2}
Z[\vect{0} \rightarrow \vect{h}_{f},\tau_{f}]
={\e}^{-\frac{1}{2\Gamma}\vect{h}_{f}.W_{N}\vect{h}_{f}}
\int {\cal D}\vect{h}\,\delta^{d}(\vect{h}_{N}-\vect{h}_{f})
{\e}^{ -\frac{1}{2}\sum_{i=0}^
{N-1}\vectg{\psi}_{i+1}.\tilde{Q}_{i}\vectg{\psi}_{i+1}}\,,
\end{equation}
where we defined the new field $\vectg{\psi}_{i+1}$ (for $0\leq i \leq N-1$) as
\begin{equation}
\label{defpsi}
\vectg{\psi}_{i+1}=\vect{h}_{i+1}-\vect{h}_{i}-\Delta \tau \tilde{Q}_{i}
^{-1}({\cal B}_{i}+\frac{W_{i}+W_{i+1}}{2})\,
\frac{\vect{h}_{i}+\vect{h}_{i+1}}{2}\,,
\end{equation}
and the measure of integration ${\cal D}\vect{h}$ as
\begin{equation}
{\cal D}\vect{h}=\prod _{i=0}^{N-1} \frac{d\vect{h}_{i+1}}{(2\pi \Gamma 
\Delta \tau )^{d/2}}\,.
\end{equation}
With the help of (\ref{defU}), the transformation (\ref{defpsi}) may be 
rewritten as
\begin{equation}
\label{lienpsih}
\vectg{\psi}_{i+1}=\vect{h}_{i+1}-\vect{h}_{i}-\left(\frac{U_{i+1}-U_{i}}{2}
\right)\left(\frac{U_{i} +U_{i+1}}{2}\right)\!^{-1}
\,(\vect{h}_{i}+\vect{h}_{i+1})\,.
\end{equation}
This relation is easily inverted by setting 
$\vect{h}_{i}=U_{i}\vectg{\zeta}_{i}$, which gives
\begin{eqnarray*}
\vectg{\psi}_{i+1}&=&\left\{\frac{U_{i}+U_{i+1}}{2}-\frac{U_{i+1}-U_{i}}{2} 
\left(\frac{U_{i}+U_{i+1}}{2}\right)\!^{-1}\frac{U_{i+1}-U_{i}}{2}\right\}
(\vectg{\zeta}_{i+1}-\vectg{\zeta}_{i})\\
&=&U_{i+1}\left(\frac{U_{i}+U_{i+1}}{2}\right)^{-1}U_{i}(\vectg{\zeta}_{i+1}-
\vectg{\zeta}_{i})=\left(\frac{U_{i}^{-1}+U_{i+1}^{-1}}{2}\right)\!^{-1}
\!(\vectg{\zeta}_{i+1}-\vectg{\zeta}_{i})\,,
\end{eqnarray*}
so that we deduce (under the assumption $\vect{h}_{0}=\vect{0}$), for
$0\leq i \leq N-1$,
\begin{equation}
\label{lienhpsi}
h_{i+1}=U_{i+1}\left[\sum_{j=0}^{i}\left(\frac{U_{j}^{-1}+U_{j+1}^{-1}}{2}
\right)\,\vectg{\psi}_{j+1}\right]\,.
\end{equation}
Since $\vect{h}_{i+1}$ is linearly related to the $\vectg{\psi}_{j+1}$'s
 of index
$j$ lower than $i$, only the diagonal blocks $U_{i+1}\left(\frac{U_{i}^{-1}
+U_{i+1}^{-1}}{2}\right)$ enter the Jacobian of the transformation and
one has~:
\begin{equation}
\label{defJacobian}
J_{N}\,\equiv \left| \frac{\partial \vect{h}_{i+1}}
{\partial \vectg{\psi}_{j+1}}\right|
=\prod_{i=0}^{N-1}\frac{\det (U_{i}+U_{i+1})}{\det 2U_{i}}\,.
\end{equation}
To enforce the boundary condition $\vect{h}_{N}=\vect{h}_{f}$ at time
$\tau_f$ in terms of the new variables $\vectg{\psi}_{i}$, we introduce the
usual integral representation of the $\delta $-function~:
\begin{equation}
\delta^{d}(\vect{h}_{f}-\vect{h}_{N})=\int\,\frac{d\vectg{\alpha}}
{(2\pi )^{d}}{\e}^{-i\vectg{\alpha}.[\vect{h}_{f}-\sum_{i=0}^{N-1}
\frac{U_{i}^{-1}+U_{i+1}^{-1}}{2}\vectg{\psi}_{i+1}]}\,.
\end{equation}
After performing the Gaussian integration over the $\vectg{\psi}_{i}$'s, one
arrives at 
\begin{equation}
\label{Z_mobile3}
Z[\vect{0} \rightarrow \vect{h}_{f},\tau_{f}]
=\prod_{i=0}^{N-1}\left\{\frac{\det (U_{i}+U_{i+1})}{\det 2U_{i}}
\times \frac{1}{\sqrt{\det \tilde{Q}_{i}}}\right\}\,
{\e}^{-\frac{1}{2\Gamma}\vect{h}_{f}.W_{N}\vect{h}_{f}}\times 
\int\,\frac{d\vectg{\alpha}}{(2\pi )^{d}}{\e}^{-i\vectg{\alpha}\cdot\vect{h}_{f}}
{\e}^{-\frac{\Delta \tau \Gamma}{2}\vectg{\alpha}.G\vectg{\alpha}}\,,
\end{equation}
where
\begin{equation}
\label{defG}
G=U_{N}\,\left[\sum_{i=0}^{N-1}\left(\frac{U_{i}^{-1}+U_{i}^{-1}}{2}\right)
\tilde{Q}_{i}^{-1}\left(\frac{^{t}U_{i}^{-1}+{^{t}U_{i+1}^{-1}}}{2}\right)
\right]\,^{t}U_{N}\,.
\end{equation}
This awkward non local operator $G$ is greatly simplified when the singular
initial condition already mentioned~: $U_{0}=\epsilon$, $W_{0}=\epsilon^
{-1}$ with $\epsilon \rightarrow 0^{+}$ is adopted. In that case, $G$ is
completely dominated by the first term of the series in the r.h.s. of
(\ref{defG}) which diverges as $\epsilon^{-1}$ and, to leading order in
$\epsilon$, one has
\begin{equation}
G\approx \frac{1}{\Delta \tau}U_{N}\left(U_{0}^{-1}W_{0}^{-1} 
{^{t}U_{0}^{-1}}\right) {^{t}U_{N}}\,.
\end{equation}
The summation over $\vectg{\alpha}$ can then be done and, since
$G^{-1}$ vanishes in the $\epsilon \rightarrow 0^{+}$  limit, one gets
\begin{equation}
\label{Z_mobile4}
 Z[\vect{0} \rightarrow \vect{h}_{f},\tau_{f}]
=\left(\frac{1}{2\pi \Gamma \Delta \tau}\right)^{d/2}
\frac{\det (U_{0}+U_{1})}{\det U_{N}}
\prod_{i=1}^{N-1}\left\{\frac{\det (U_{i}+U_{i+1})}{\det 2U_{i}}
\times \frac{1}{\sqrt{\det \tilde{Q}_{i}}}\right\}\,
{\e}^{-\frac{1}{2\Gamma}\vect{h}_{f}.W_{N}\vect{h}_{f}}\,.
\end{equation}

Note that all the manipulations presented in this Appendix 
were devoid of any approximation. 
It is finally a straightforward matter (details will be skipped here), to 
check that in the $\Delta \tau \rightarrow 0$ limit, the infinite
product in front of the exponential in (\ref{Z_mobile4}) reduces to
$\left(\frac{1}{2\pi \Gamma \Delta \tau}\right)^{d/2}{1\over \sqrt{\det U_{N}}}$, making thereby (\ref{Z_mobile4}) identical to the result
(\ref{Z_mobile}) quoted in the text.

\section{}
\label{AppendixIII}

We consider the following quadratic action~:
\begin{equation}
\label{C1}
S_{2}[x_{n}]=\frac{1}{2}\sum_{n=0}^{d-1}\,\int_{0}^{\tau } d\tau \,\frac{\dot{x}_{n}^{2}}{a_{n}(\tau )},
\end{equation}
where, in order to mimic the stochastic models studied in this paper, the
variance $a_{n}(\tau )$ evolves on each shell as
\begin{equation}
\label{C2}
a_{n}(\tau )=(\vect{b}^{0}\cdot\vect{b}^{0})(C_{n-2}^{0}C_{n-1}^{0})^{2},
\end{equation}
i.e., as the variance of noise, $(\vect{b}^{0}\cdot\vect{b}^{0}) B_{nn}$, for the model (ii). In (\ref{C2}), $\vect{b}^{0}(\tau )$ may be thought of as the
deterministic self-similar solution, and $a_{n}(\tau )$ can therefore be
cast into the form 
\begin{equation}
\label{C3}
a_{n}(\tau )={\e}^{2A_{0}\tau } \tilde{a}(\tau -nT_{0}),
\end{equation}
where the function $\tilde{a}(\tau )$ satisfies
\begin{equation}
\label{C4}
\tilde{a}(\tau+dT_{0})=\tilde{a}(\tau ),
\end{equation}
because of the periodicity of the shell lattice. Let us assume that the instanton is centered around the shell of index $n=0$ at time $\tau =0$.
At its leading edge ($n>0$),  $C_{n}$ decreases very abruptly as
$\exp -cr^{n}$, with r=$(\sqrt{5}-1)/2$ and $c$ a constant of order 1. This
essential singularity comes from the necessity of balancing $\frac{dC_{n}}
{d\tau }$ with the dominant term $Q^{2}(1-\epsilon )C_{n-2}C_{n-1}$ of
 the non-linear kernel of the GOY model in this range of scales. In the trail
 of the instanton ($n<0$), one has a much smoother behaviour
$C_{n} \sim Q^{nz_{0}}\equiv {\e}^{nA_{0}T_{0}}$. When the shell lattice
is periodized, the leading edge and the tail of the instanton have to be glued together and the locus of matching, as well as the residual amplitude of $C_{n}$ at that place, will be imposed by the side supporting the slowest
 variations of $C_{n}$. We conclude that in a cyclic chain
containing $d$ shells, most of them reside in the exponential tail of the
instanton, so that we may write (again under the hypothesis of an instanton initially centered around the origin $n=0$ and with a shell index
$n$ defined between 0 and $d-1$)
\begin{equation}
\label{C5}
a_{n}(0)\sim {\e}^{4(n-d)A_{0}T_{0}},  
\end{equation}
Thus, the range of values spanned by the function $\tilde{a}$ is very
large and scales with the total number of shells like ${\e}^{4dA_{0}T_{0}}$.
 For
shells on the exponential ramp, $a_{n}(\tau )$ first decreases exponentially
in time like ${\e}^{-2A_{0}\tau }$ (because $C_{n}$ decreases like 
${\e}^{-A_{0}\tau }$ in this region) and goes by a sharp maximum 
of order ${\e}^{2nA_{0}T_{0}}$ at the time $\tau_{n} \sim nT_{0}$ when the center of the instanton reaches the corresponding shell. Then it starts again to decrease exponentially. \\
Having understood these basic dynamical features, we can compute the
matrices $U$ and $W$ introduced in subsection \ref{fluctformal} for the
 quadratic action
given by (\ref{C1}). To make contact with the normalized field $\vect{h}$ 
used in the real problem, we switch from the variable $x_{n}$ to the 
variable $y_{n}=x_{n}/\sqrt{a_{n}}$. This transforms the original action into
\begin{equation}
\label{C6}
S_{2}[y_{n}]=\frac{1}{2}\sum_{n=0}^{d-1}\,\int_{0}^{\tau } d\tau \,
 (\dot{y}_{n}+\frac{1}{2}\frac{\dot{a}_{n}}{a_{n}}y_{n})^{2}.
\end{equation}
Since there is no inter-shell coupling, the matrices $U$ and $W$ are 
diagonal in the shell index. The extremization of $S[y_{n}]$
with respect to $y_{n}$ leads to the couple of first-order differential
equations 
\begin{eqnarray}
\label{C7}
p_{n}&=&\dot{y}_{n}+\frac{1}{2}\frac{\dot{a}_{n}}{a_{n}}y_{n}, \\
\dot{p}_{n}&=&\frac{1}{2}\frac{\dot{a}_{n}}{a_{n}}p_{n}.
\label{C8}
\end{eqnarray}
One has simply $U_{nn}=y_{n}$ and $W_{nn}U_{nn}=p_{n}$. The solution of
Eqs. (\ref{C7}) and (\ref{C8}) under the initial conditions $y_{n}(0)=0$ and
$p_{n}(0)=1$ is
\begin{eqnarray}
\label{C9}
p_{n}(\tau )&=&\frac{\sqrt{a_{n}(\tau )}}{\sqrt{a_{n}(0)}},\\
y_{n}(\tau )&=& \frac{1}{\sqrt{a_{n}(0)}}\frac{1}{\sqrt{a_{n}(\tau )}}
\int _{0}^{\tau }a_{n}(\tau ') {d\tau '}.
\label{C10}
\end{eqnarray}
As long as $\tau < \tau_{n}=nT_{0}$, the instanton has not passed through the shell of index $n$ and the integral in the right hand side of (\ref{C10}) is dominated by the neighborhood of the lower bound $\tau =0$. We
deduce that for indices $n> \tau /T_{0}$
\begin{eqnarray}
\label{C11}
U_{nn}(\tau ) &=&\frac{{\e}^{A_{0}\tau }}{2A_{0}}, \\
W_{nn}(\tau )&=&2A_{0}\,{\e}^{-2A_{0}\tau }.
\label{C12}
\end{eqnarray}
By contrast, when $\tau $ gets larger than $\tau_{n}$ (by some units of
 time $T_{0}$), the integral in the right hand side of (\ref{C10}) is 
dominated by the neighborhood of the time $\tau_{n}$ where $a_{n}$
takes its maximal value. We get 
\begin{eqnarray*}
y_{n}(\tau ) &\sim & I\,\sqrt{\frac{a_{n}(\tau_{n})}{a_{n}(0)}}
{\e}^{A_{0}(\tau -\tau_{n})},\\
p_{n}(\tau )&\sim &\sqrt{\frac{a_{n}(\tau_{n})}{a_{n}(0)}}
{\e}^{-A_{0}(\tau -\tau_{n})},
\end{eqnarray*}
where $I$ is a number of order 1. It follows that for indices 
$n< \tau /T_{0}$,
\begin{eqnarray}
\label{C13}
U_{nn}(\tau )&\sim & I\,{\e}^{2(d-n)A_{0}T_{0}}\,{\e}^{A_{0}\tau }, \\
W_{nn}(\tau )&\sim &I^{-1}\,{\e}^{2nA_{0}T_{0}}\,{\e}^{-2A_{0}\tau },
\label{C14}
\end{eqnarray}
where we used the estimate (\ref{C5}) for $a_{n}(0)$. Since 
$W_{nn}U_{nn} \sim {\e}^{-A_{0}\tau }$ for $n>\tau /T_{0}$ and 
$\sim {\e}^{2dA_{0}T_{0}} {\e}^{-A_{0}\tau}$ for $n<\tau /T_{0}$, we
conclude that $\det WU$ increases exponentially like ${\e}^{dA_{0}\tau }$. But this property is not shared by $\det U$ or $\det W$ considered 
individually. Indeed, since the number of shells crossed by the instanton
 increases linearly in time, eqs. (\ref{C11}) and (\ref{C13}) show that
\begin{equation}
\label{C15}
 \det U \sim \left(\frac{I}{2A_{0}}\right)^{\tau /T_{0}}{\e}^{3A_{0}d\tau}
{\e}^{-A_{0}\frac{\tau ^{2}}{T_{0}}},
\end{equation}
while from eqs. (\ref{C12}) and (\ref{C14}), 
\begin{equation}
\label{C16}
 \det W \sim \left(\frac{I}{2A_{0}}\right)^{-\tau /T_{0}}
{\e}^{-2A_{0}d\tau}{\e}^{+A_{0}\frac{\tau ^{2}}{T_{0}}}.
\end{equation}
This argument captures apparently a good part of the physics of 
fluctuations around a moving self-similar system, though badly treating
hybridization effects between neighbouring shells. It also explains how
large (resp. small) numbers are generated in the spectrum of the matrix
$U$ (resp. $W$) and why in practice one does not have much
 freedom in the choice of the total number of shells $d$.

\newpage
 
\newpage

\newpage

\newpage

\newpage

\newpage

\newpage



\begin{references}
\bibitem[21]{rem1} Actually, this change of variable is not strictly necessary.
It may be shown to roughly ``square root'' the range of values spanned by
the coefficients of matrices $U$ and $W$ introduced in subsection
 \ref{fluctformal}. The ensuing reduction of numerical stiffness allows
computations at larger sizes of the shell lattice. However, the switch 
from $\vectd{\delta}{b}$ to $\vect{h}$ gets totally impracticable when the
matrix $B$ cannot be easily inverted.
\end{references}
\end{document}